%% file: ms.tex
\documentclass{emulateapj}
\usepackage{graphicx}
\usepackage{natbib}
\slugcomment{To appear in ApJ., XX.}
\shorttitle{Clues to SN2009ip from LCOGT}
\shortauthors{Graham et al.}
\begin{document} 
\title{Clues To The Nature of SN\,2009ip from \\ Photometric and Spectroscopic Evolution to Late Times}
\author{
M.L. Graham\altaffilmark{1},
D.J. Sand\altaffilmark{2},
S. Valenti\altaffilmark{3,4},
D.A. Howell\altaffilmark{3,4},
J. Parrent\altaffilmark{3,5}, \\
M. Halford\altaffilmark{6},
D. Zaritsky\altaffilmark{6},
F. Bianco\altaffilmark{7}, 
A. Rest\altaffilmark{8},
B. Dilday\altaffilmark{9}}

\altaffiltext{1}{Astronomy Department, University of California at Berkeley, Berkeley CA 94720}
\altaffiltext{2}{Physics Department, Texas Tech University, Lubbock, TX 79409}
\altaffiltext{3}{Las Cumbres Observatory Global Telescope Network, Goleta CA 93117}
\altaffiltext{4}{Physics Department, University of California at Santa Barbara, Santa Barbara CA 93106}
\altaffiltext{5}{Harvard-Smithsonian Center for Astrophysics, 60 Garden St, Cambridge, MA 02138}
\altaffiltext{6}{Astronomy Department, University of Arizona, Tucson, AZ 85721}
\altaffiltext{7}{Department of Physics, New York University, 4 Washington Place, New York, NY 10003}
\altaffiltext{8}{Space Telescope Science Institute, 3700 San Martin Dr., Baltimore MD 21218}
\altaffiltext{9}{North Idaho College, 1000 W. Garden Ave, Coeur d'Alene, ID 83814}

\begin{abstract}

We present time series photometric and spectroscopic data for the transient SN\,2009ip from the start of its outburst in September 2012 until November 2013. This data was collected primarily with the new robotic capabilities of the Las Cumbres Observatory Global Telescope Network, a specialized facility for time domain astrophysics, and includes supporting high-resolution spectroscopy from the Southern Astrophysical Research Telescope, Kitt Peak National Observatory, and Gemini Observatory. Based on our nightly photometric monitoring, we interpret the strength and timing of fluctuations in the light curve as interactions between fast-moving ejecta and an inhomogeneous CSM produced by past eruptions of this massive luminous blue variable (LBV) star. Our time series of spectroscopy in 2012 reveals that, as the continuum and narrow H$\alpha$ flux from CSM interactions declines, the broad component of H$\alpha$ persists with SN-like velocities that are not typically seen in LBVs or SN Impostor events. At late times we find that SN\,2009ip continues to decline slowly, at $\lesssim0.01$ magnitudes per day, with small fluctuations in slope similar to Type IIn supernovae (SNe\,IIn) or SN impostors, but no further LBV-like activity. The late time spectrum features broad calcium lines similar to both late time SNe and SN Impostors. In general, we find that the photometric and spectroscopic evolution of SN\,2009ip is more similar to SNe\,IIn than either continued eruptions of an LBV star or SN Impostors. In this context, we discuss the implications for episodic mass loss during the late stages of massive star evolution.

\end{abstract}


\section{Introduction} \label{sec:intro}

Type IIn supernovae (SN\,IIn) are characterized by bright optical emission caused when explosion ejecta collides with nearby circumstellar material (CSM). This reveals that the progenitor star experienced a significant amount of mass loss prior to core collapse. The physical mechanism driving the mass loss, and the relative contributions to the CSM from winds and episodic eruptions, are not yet well constrained. In the final stages of massive star evolution, the star undergoes a rapid succession of burning phases, and exhibits variability during its mass loss episodes. The distribution of material in the CSM can be derived from spectroscopic and photometric fluctuations of SNe\,IIn which, if the star was monitored for variability in previous years, allows for a connection between pre-explosion mass loss episodes and the material they produce. Luckily, the transient SN\,2009ip affords us this rare opportunity.

SN\,2009ip was first identified as a photometric transient near host galaxy NGC 7259 by the Chilean Automatic Supernova Search (CHASE; Maza et al. 2009), and given the designation `SN' without spectroscopic confirmation. Archival imaging of Hubble Space Telescope images of the progenitor star of SN\,2009ip indicated it is very massive, $M=$ 50--80 $\rm M_{\odot}$ \citep{Smith2010}. Analysis of archival data found a coincident variable source and suggested the transient may in fact be a luminous blue variable (LBV) or cataclysmic variable \citep{Miller2009}. The LBV hypothesis was quickly confirmed with spectroscopy \citep{Berger2009}, which showed SN\,2009ip was similar to the group of ``SN imposters" that look like Type IIn supernovae (e.g. Van Dyk et al. 2000). After peak brightness in 2009, the transient fell but then rose again in a manner similar to eta Carinae \citep{Li2009}, and experienced a second outburst in mid-2010 \citep{Drake2010}. Extensive analysis and discussion of the evidence for the 2009 and 2010 activity of SN\,2009ip as an LBV eruption are provided by Smith et al. (2010) and Foley et al. (2011). In August 2012, SN\,2009ip re-brightened to record luminosity \citep{Drake2012}, maintaining an LBV-like spectrum \citep{Foley2012}, but by mid-September 2012 the spectral lines had broadened and shifted, indicating material at velocities up to 13,000 $\rm km\ s^{-1}$, and showed P-Cygni profiles typical of core collapse supernova explosions \citep{Smith2012}.

\begin{figure}[t]
\begin{centering}
\includegraphics[trim=0.7cm 0.0cm 0.0cm 0.0cm, clip, width=8.7cm]{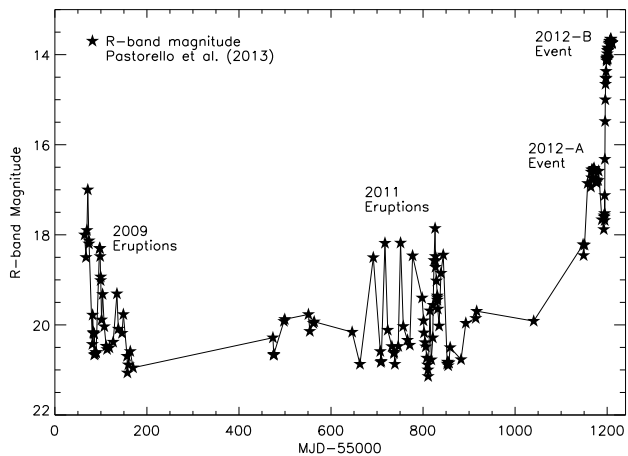}
\caption{Past eruptions of SN\,2009ip as presented in Pastorello et al. (2013), and reproduced here for context. Lines connect the data points to guide the eye, but do not represent continuous photometric monitoring. \label{fig:pastorello}}
\end{centering}
\end{figure}

An analysis of spectroscopic and photometric data during the outburst of SN\,2009ip from September to December of 2012 was presented in Mauerhan et al. (2013, hereafter JM13). They report Balmer lines with broad P-Cygni profiles and high absorption component velocities characteristic of core collapse supernovae, and completely unprecedented by any previous LBV eruption. JM13 observed SN\,2009ip to fade, and then brighten to $M_R = -17.5$. The outbursts before and after this temporary fading are referred to as the 2012-A and 2012-B events (see Figure \ref{fig:pastorello}). At the time of re-brightening the broad component of the Balmer lines disappeared, and only the narrow Lorentzian component remained. SN\,2009ip then resembled a Type IIn supernovae, in which emission is dominated by the interaction of SN ejecta with the circumstellar medium (CSM). Based on their data, JM13 concluded that the 2012-A event of SN\,2009ip was the core collapse of an LBV star during an outburst, and that the 2012-B event was caused by the fast-moving ejecta interacting with the CSM. This conclusion was supported by Prieto et al. (2013), who present a densely time-sampled optical light curve from the start of the 2012-B event and find that the rapid rise and bolometric luminosity are similar to other SNe\,IIn. 

The interpretation of SN\,2009ip as a SN\,IIn was challenged by Pastorello et al. (2013, hereafter AP13) who present three years of monitoring of SN\,2009ip; we have included their photometry for context in Figure \ref{fig:pastorello}. They find that spectra from September 24 2011 also exhibit material at velocity $\sim13000$ $\rm km\ s^{-1}$, showing for the first time that non-terminal explosions can create such fast ejecta, and calling into question the supernova diagnosis for SN\,2009ip. Instead, AP13 suggest the 2012 activity of SN\,2009ip was caused by a pair-instability event during 2012-A, with the 2012-B event caused by ejecta material colliding with the CSM. The scenario of AP13 is fundamentally different from JM13 because the the star is not destroyed . This hypothesis is supported by the analysis of Fraser et al. (2013, hereafter MF13), who find a very low upper limit of $0.02 M_{\odot}$ on the mass of any synthesized material (e.g. $\rm ^{56}Ni$). A similar inference is made by Margutti et al. (2013, hereafter RM13), who analyze extensive multi-wavelength observations for SN\,2009ip's 2012 activity, and conclude that the events were not caused by a core collapse but that the true physical mechanism behind the explosive ejection in the LBV's envelope could not, at that time, be fully distinguished. A full analysis including late-time observations of SN\,2009ip by Smith et al. (2013, hereafter NS13) finds that the broad component of H$\alpha$, first announced in Smith et al. (2012), persisted throughout the 2012-B event, indicating a significantly more energetic explosion has occurred than considered by RM13. They also find that the late-time spectrum of SN\,2009ip no longer resembles an LBV. Combined, the observations of NS13 lead them firmly to the conclusion that a terminal supernova explosion has occurred.

This paper presents new, densely time-sampled photometry and spectroscopy for SN\,2009ip from the The Las Cumbres Observatory Global Telescope Network (LCOGT.net), and other facilities at which we were observing classically at the time. LCOGT is a new system of robotic telescopes dedicated to time domain astrophysics. Telescopes of 1 and 2 meter diameter are distributed at five sites around the world, and observation requests are optimized and automatically executed by a real-time adaptive scheduler. Instrumentation includes identical imagers with a full suite of filters on all telescopes, and one robotic spectrograph (FLOYDS) on each of the two 2 meter Faulkes Telescopes. The distributed aspect of LCOGT make it flexible, responsive, and unhindered by weather or sunrise. New astrophysical transients are followed as soon as possible for classification, and can be reliably monitored on any desired time scale. The LCOGT Network, presented in depth by Brown et al. (2013), currently features the following facilities: McDonald Observatory, Texas USA (one 1m); Haleakala Observatory, Hawaii (one 2m+FLOYDS); Sutherland, South Africa (three 1m); Cerro Tololo, Chile (three 1m); and Siding Spring, Australia (two 1m and one 2m+FLOYDS).

The 2012--2013 activity of SN\,2009ip is one of the first astrophysical transients followed with the LCOGT 1m network, and is a great example of the science potential from the combination of reactive follow-up and long term monitoring with a network that includes photometric and spectroscopic capabilities. The quality, quantity, and extent of these data allow us to provide new insight to the controversial physical interpretation of SN\,2009ip, especially during it's extended slow decline. In \S~\ref{sec:obs} we present the observations; in \S~\ref{sec:LC} and \ref{sec:spec} we interpret the photometry and spectroscopy respectively, and discuss the results in context with the ongoing debate about the true physical nature of SN\,2009ip before concluding in \S~\ref{sec:con}. A flat cosmology with $\Omega_M=0.27$ and $\Omega_{\Lambda}=0.73$ is assumed throughout.

\section{Observations} \label{sec:obs}

Here we present the photometric and spectroscopic observations of SN\,2009ip obtained with the imaging cameras and FLOYDS spectrograph on the LCOGT 1 and 2 meter telescopes, the Goodman Spectrograph on the 4 meter SOAR Telescope at Cerro Pachon, the RC Spectrograph on the Mayall 4 meter telescope at Kitt Peak National Observatory, and the Gemini Multi-Object Spectrograph on the 8m Gemini South telescope.

\subsection{LCOGT 1.0m and 2.0 Photometry}

\begin{figure}[t]
\begin{center}
\includegraphics[width=8cm]{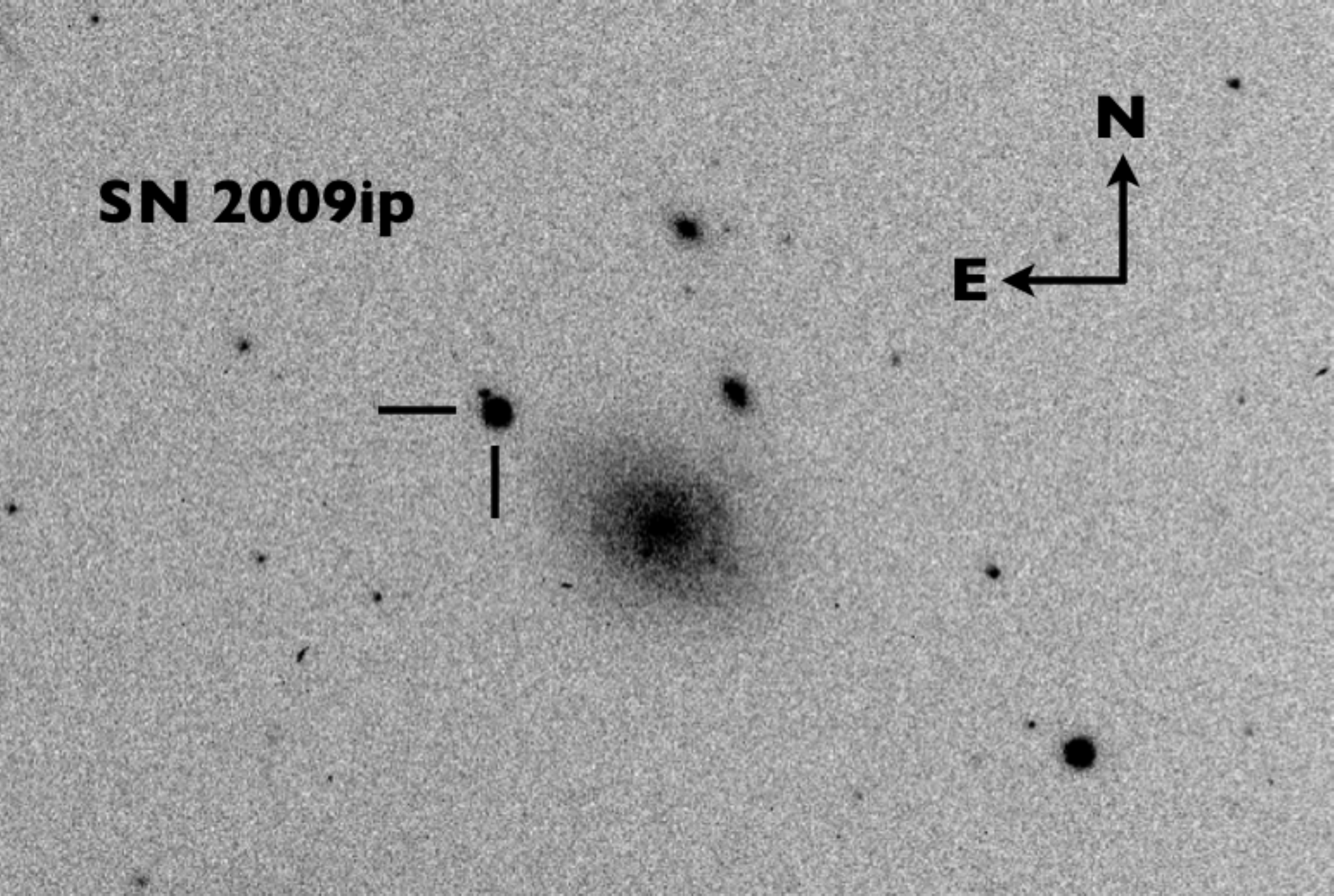}
\caption{Image of SN\,2009ip taken on 2012-10-21 in $r^{\prime}$ with one of the LCOGT 1.0 meter telescopes at CTIO. \label{fig:image}}
\end{center}
\end{figure}

\begin{figure}[t]
\epsscale{1.2}
\plotone{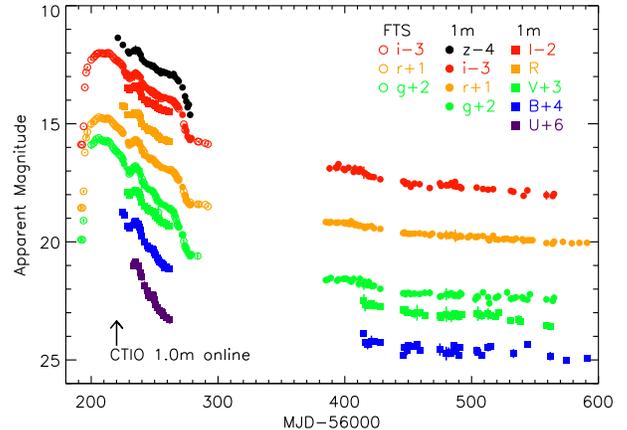}
\caption{The full extent of all LCOGT photometry for SN\,2009ip, from the start of the 2012-B event to November 2013, in all filters. Observations were collected in $gri$ filters with the 2.0m FTN until the 1.0m telescopes at CTIO became available, as marked. At late times, the 1.0m photometry was collected with both CTIO and SAAO telescopes. \label{fig:lc_full}}
\end{figure}

Photometric monitoring of SN\,2009ip began on 2012-09-22 with the Spectral camera on Faulkes Telescope South (Siding Spring Observatory, Australia) in filters $g^{\prime}r^{\prime}i^{\prime}$. The Spectral camera is a 4096 $\times$ 4096 pixel Fairchild CCD with a 10.5$\arcmin$\ field of view. As Faulkes South monitored the rise of SN\,2009ip, LCOGT's first three southern 1m telescopes arrived on site at Cerro Tololo International Observatory in Chile. Photometric monitoring of SN\,2009ip from CTIO began when the telescopes were operational, about one month later on UT 2012-10-21, with the first generation deployment camera: SBIG CCDs with a 15\arcmin field of view. In Figure \ref{fig:image} we show one of the first LCOGT $r^{\prime}$ images of SN\,2009ip, and in Figure \ref{fig:lc_full} we show the full LCOGT light curve, indicating the point at which the CTIO 1.0m came online. We monitored SN\,2009ip until the end of 2012 with as close to daily cadence as possible, in Johnson-Cousins filters $UBVRI$, Sloan filters $g^{\prime}r^{\prime}i^{\prime}$, and Pan-STARRS z-short. When SN\,2009ip was again accessible in May 2013, we monitored it with the 1.0m network, including the new site at the South African Astronomical Observatory (SAAO), with near daily cadence in $BVg^{\prime}r^{\prime}i^{\prime}$. All images have been processed through the LCOGT automatic reduction pipelines. We used observations of both SN\,2009ip and SDSS Landolt fields on photometric nights to build a catalog of standard stars in the field of SN\,2009ip, from which we calibrated the photometry. In the process, $g^{\prime}r^{\prime}i^{\prime}$ have been converted to SDSS $gri$. The photometry has been corrected for Milky Way extinction only where explicitly mentioned; otherwise, calibrated observed apparent magnitudes are presented. We present our photometry from the LCOGT 2.0 Faulkes Telescope and CTIO 1m telescopes in Table \ref{tab:allphot}.

\subsection{LCOGT FLOYDS Spectroscopy}

Spectroscopic monitoring of SN\,2009ip began on UT 2012-09-25 with the new FLOYDS robotic spectrograph on the 2.0m Faulkes Telescope South (FTS) at Siding Spring Observatory in Australia. A twin spectrograph is located at the Faulkes Telescope North (FTN) on Haleakala, Hawaii. These robotic spectrographs have low resolution and were specifically designed for supernovae. The design is a simple folded Schmidt camera, with a double-pass prism and reflective grating. FLOYDS uses a standard reflection grating (235 l/mm) as the primary disperser, with a cross-dispersed prism to image the first (5400--10000 \AA) and second (3200--5700 \AA) order light onto the CCD. This provides an overall wavelength coverage from 3200 \AA\ to 10000 \AA\ in a single exposure. Slit widths range from 0.9\arcsec\ to 6.0\arcsec, but only the 2.0\arcsec\ slit was used in this work. The resolution in the red is R $\sim$ 400, or 16.5 \AA\ per resolution element, and in the blue is R $\sim$ 690, or 8.25 \AA\ per resolution element. The LCOGT FLOYDS spectrographs, and the automatic pipeline with which these spectra were reduced, will be presented in Sand et al. (in prep). FLOYDS-FTS achieved first light on May 7, 2012, and was still in commissioning when SN\,2009ip was observed. In particular, the robotic target acquisition software was still being refined. As a result, some of the spectra suffer more from flux mis-calibration and residual fringing (e.g. 2012-10-04) than FLOYDS spectra being obtained and reduced in the present day. All FLOYDS spectra were flux calibrated with LCOGT photometry.

\subsection{SOAR Goodman Spectroscopy}

Spectroscopy of SN\,2009ip was obtained on UT 2012-09-22 with the Goodman High Throughput Spectrograph \citep{Clemens2004} on the 4 meter Southern Astrophysics Research Telescope (SOAR) at Cerro Pachon in Chile. This spectrograph has a 4k $\times$ 4k Fairfield imaging CCD. We used the long slit with a 0.84\arcsec\ width, the 600 lines per millimeter grating, and the GG385 order-blocking filter in the ``mid" wavelength option for 4350-7020 \AA\  coverage. These settings result in a readnoise of 3.99 electrons, a gain of 2.06 electrons per ADU, and resolution of R $\sim$ 1500 at 5500 \AA. Three 600 second exposures of SN\,2009ip were reduced with standard IRAF procedures, corrected for extinction, flux calibrated with standard star LTT2145, and median combined.

\subsection{Mayall Spectroscopy}

Spectroscopy of SN\,2009ip was obtained nightly from UT 2012-10-11 to 2012-10-15 with the Ritchey-Chretien (RC) Spectrograph on the 4.0 meter Mayall Telescope at Kitt Peak National Observatory. These spectra were also presented in RM13, but have been reduced and calibrated independently for this work. During the first two nights, lower resolution spectra were obtained, but because the spectra from each of these five nights are very similar, we will only present and discuss the higher resolution spectra obtained on the last three nights. For these, the higher resolution grating KPC-22B (632 lines/mm) was used in the first order, giving a spectral resolution of $\sim$2.7 \AA \ and coverage $\lambda=$ 5500--8000 \AA. All exposure times were 1200 seconds, with five, three, and four exposures taken on 2012-10-13, 14, and 15 respectively. All Mayall spectroscopy has been flux calibrated using LCOGT photometry.

\subsection{Gemini Spectroscopy}

Spectroscopy of SN\,2009ip was obtained on UT 2013-06-12 with the Gemini Multi-Object Spectrograph on the 8.0m Gemini South Telescope at Cerro Pachon, Chile (PI Howell, program GS-2012A-Q-62). We used a 1.0\arcsec \ longslit and an exposure time of 900 seconds for both the B600 and R400 gratings, with the OG515 order blocking filter on the red side. This gave spectral resolutions of 2.7 and 4.0 \AA \  on the blue and red sides respectively, and coverage over $\lambda=$ 3500--9500 \AA. These spectra were reduced and combined with standard procedures.

\begin{figure}[t]
\epsscale{1.2}
\plotone{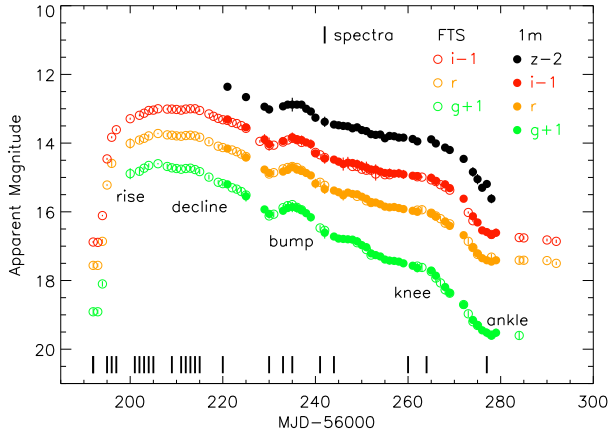}
\caption{The light curve of SN\,2009ip during the 2012-B event from FTS and CTIO in Sloan filters $gri$ and $z$-short, with magnitude offsets added for clarity. For the purpose of discussion, descriptive eras have been defined based on light curve features: rise, decline, bump, knee, and ankle. The short black lines along the bottom of the figure mark the dates of spectroscopy. \label{fig:lc_griz}}
\end{figure}

\begin{figure}[t]
\epsscale{1.2}
\plotone{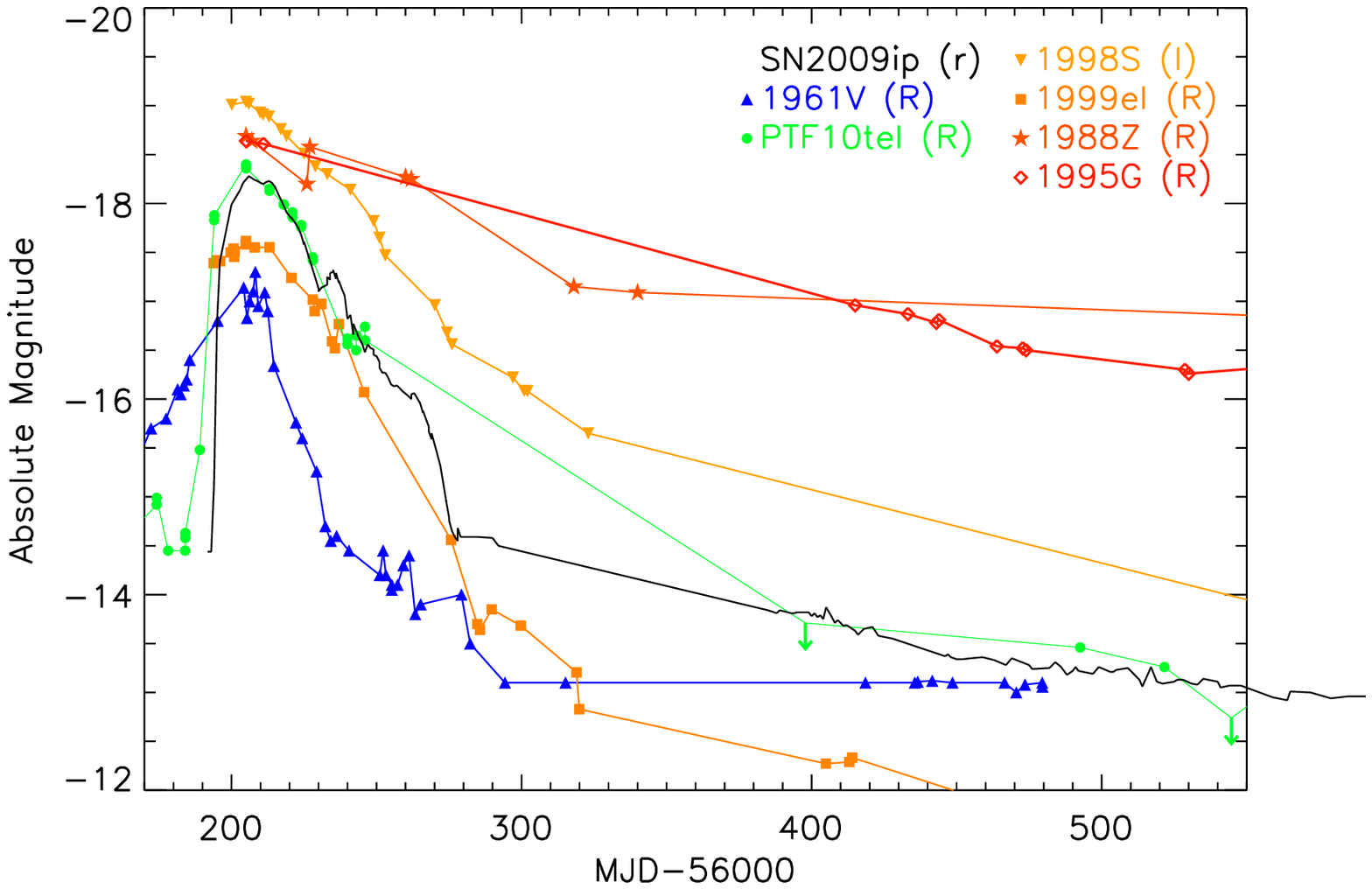}
\caption{Here we compare our light curve of SN\,2009ip (black line) to other explosions, with phases shifted to match the time of peak brightness for SN\,2009ip. Blue triangles are the potential SN impostor event 1961V \citep{Bertola1963}; green circles are PTF10tel (aka SN2010mc; Ofek et al. 2013a); orange inverted triangles are SN\,IIn 1998S \citep{Fassia2000}; orange-red squares are SN\,IIn 1999el \citep{dicarlo2002}; red-orange stars are SN\,IIn 1988Z (Turatto et al. 1993; Stathakis \& Sadler 1991); and red diamonds are SN\,IIn 1995G \citep{Pastorello2002}. \label{fig:lc_comp}}
\end{figure}

\section{The Light Curve of SN\,2009ip}\label{sec:LC}

Photometric monitoring of SN\,2009ip with the LCOGT network began on 2012-09-22, in between the 2012-A and 2012-B events. We registered no change in magnitude on the next day, 2012-09-23. After this the transient increased rapidly in brightness, as seen in Figure \ref{fig:lc_griz}. We chose a nightly cadence, but point out that faster cadences are possible at the LCOGT network, down to minute cadences with conventional cameras, and even faster with high speed photometers if the object is bright. Prieto et al. (2013) present an analysis based on $\sim$hourly observations of the 2012-B rise. They find a very rapid brightening on 2012-09-24 (0.5 magnitudes over 6 hours), consistent with $L\propto t^2$; this rapid rise is followed by a slower rise to peak that we see clearly in Figure \ref{fig:lc_griz}.  

We find the light curve reaches its plateau-like peak on 2012-10-06 at $m_g=13.6$, $m_r=13.7$, and $m_i=14.0$ magnitudes. For this work we have used the recession velocity for NGC 7259 from the NASA Extragalactic Database (NED\footnote{http://ned.ipac.caltech.edu/}), $v=$ 1782 $\rm km\ s^{-1}$ \citep{HIPASS}. Under our assumed cosmology, this is a distance of 25.5 Mpc, and a distance modulus of $\mu=$32.0. At the coordinates of SN\,2009ip, Milky Way extinction is $A_g=0.064$, $A_r=0.044$, $A_i=0.033$ \citep{Schlegel1998}, and we assume no contribution from host extinction given that SN\,2009ip is in the outskirts of NGC 7259. Accordingly, we find that SN\,2009ip peaked at $M_g=-18.5$, $M_r=-18.3$, and $M_i=-18.0$ magnitudes. These results agree well with published observations (e.g. JM13, Prieto et al. 2013, and MF13 use $\mu = $31.55 and find $M\sim$ -18 magnitudes; RM13 use a distance modulus of 32.05 and quote $M_R\sim-18$). 

The peak magnitude of SN\,2009ip is brighter than the typical peak of Type II-Plateau (SNe\,IIP), the most common type of core collapse explosion, that typically reach $R\sim-16$ to $-17$ and only a few reach $R\sim-18$ \citep{Li2011}. For SNe\,IIn the continuum emission is dominated by CSM interaction, the amount of which varies between individual events and causes heterogeneity in peak magnitudes for this class \citep{Li2011}. A peak magnitude of $M_r=-18.3$ is not unusual for SNe\,IIn, as we show in Figure \ref{fig:lc_comp}, and as also shown by MF13 and RM13. Historically, SN\,2009ip's brightest recorded apparent magnitude was $\rm m_R\sim18$, or $\rm M_R\sim-14.5$ (e.g. 2009-08-30, AP13); relative to its past, the 2012 events of SN\,2009ip are of an unprecedented luminosity. In general, the total change in magnitude and the smoothness of the light curve are quite distinct from the usual activity of an LBV. In these qualities, SN\,2009ip resembles a SN Impostor such as 1961V (blue triangles, Figure \ref{fig:lc_comp}). However, SN\,2009ip is significantly brighter than the typical brightest SN Impostor, $M\sim -14$; 1961V is an outlier in this respect. In particular, 1961V is so much like a real SN that new data has been acquired in recent years and the debate still continues (e.g. Kochanek, Szczygiel, \& Stanek 2011; Van Dyk \& Matheson 2012). Furthermore, SN Impostors are not typically associated with stars as massive as the known progenitor of SN\,2009ip. Interestingly, RM13 and NS13 present further evidence that SN\,2009ip is not an entirely unique event: they find a photometric sibling in SN\,IIn 2010mc, whose light curve is nearly identical to the 2012-A and 2012-B events of SN\,2009ip (green circles in Figure \ref{fig:lc_comp}; Ofek et al. 2013a).

The 2012-B event of SN\,2009ip is characterized by bumpy features indicative of irregularities in the CSM created by repeated mass loss episodes during past LBV outbursts. It was speculated that the plateau seen immediately before SN\,2009ip set was a return to post-outburst brightness, marking the end of an outburst in which the progenitor star did not explode. However, when SN\,2009ip was again accessible in April 2012 we found it still declining steadily.

In the following sections we leverage the daily cadenced multi-band photometry from LCOGT into a detailed analysis of the light curve features, and what they reveal about the physical nature of SN\,2009ip. We discuss the observed CSM interaction during the 2012-B event as probe of past mass loss episodes in \S~\ref{sec:LC_CSM}; the main light curve ``bump" era in greater detail in \S~\ref{sec:LC_bump}; the color evolution and a blackbody interpretation for the emitting material in \S~\ref{sec:LC_color}; our derivation of bolometric luminosity and total event energetics in \S~\ref{sec:LC_bolo}; and finally the power source of the late-time decline in \S~\ref{sec:LC_late}.

\begin{figure}[t]
\epsscale{1.2}
\plotone{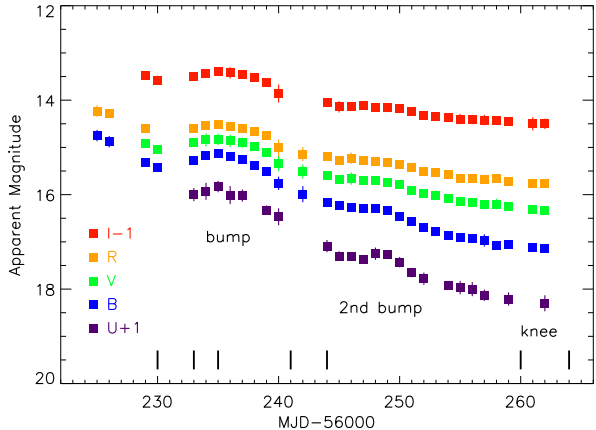}
\caption{The light curve of SN\,2009ip during the 2012-B event from CTIO in Johnson-Cousins $UBVRI$ filters, with magnitude offsets added for clarity. The ``bump" defined in Figure \ref{fig:lc_griz} for $griz$ data is also seen to start on MJD=233, but coverage in $UBVRI$ ends before the ``knee".  The short black lines mark the dates of spectroscopy. \label{fig:lc_UBVRI}}
\end{figure}

\begin{figure}[t]
\epsscale{1.2}
\plotone{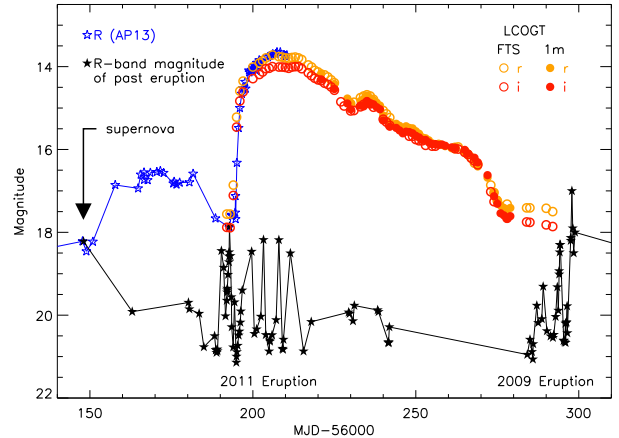}
\caption{The major re-brightening of the 2012-B event of SN\,2009ip coincides with the expected time of interaction between $v\sim500$ $\rm km\ s^{-1}$ material from past eruptions and $v\sim4500$ $\rm km\ s^{-1}$ ejecta from a supernova at the start of the 2012-A event (arrow). Black stars are photometry from past eruptions (AP13) plotted in reverse, at the expected date of interaction, as described in the text. Blue stars are $R$-band photometry for the 2012-A and B events of SN\,2009ip from AP13, and orange and red circles are LCOGT $g$ and $r$ band photometry for the 2012-B event. \label{fig:lc_past}}
\end{figure}

\subsection{CSM Interaction as a Probe of Past Eruptions}\label{sec:LC_CSM}

Far from a smooth procession, the light curve of SN\,2009ip is more of a roller coaster ride. The photometric evolution is characterized by bumpy features that occur with greater strength in bluer bands, as shown in Figures \ref{fig:lc_griz} and \ref{fig:lc_UBVRI}.  For ease of discussion, in Figure \ref{fig:lc_griz} we label five main ``eras" of the light curve: rise, decline, bump, knee, and ankle. Peak brightness is reached on modified Julian date (MJD) = 56206 (2012-10-06), but the decline does not begin in earnest until MJD = 56215 (2012-10-15). The light curve shows the obvious ``bump'' in all bands starting on MJD $\sim$ 56232 (2012-11-01), which is discussed further in Section \ref{sec:LC_bump}. A more subtle bump follows, seen most clearly in bands $g$, $B$, and $U$ on MJD = 56248 (2012-11-17). The ``knee'' in the light curve starts on MJD = 56264 (2012-12-02), after which the $g$ and $i$-bands drop more quickly than the $r$, until the ``ankle'' is reached on MJD = 56278 (2012-12-16) and the light curve flattens out. These features have been discussed in other publications, and are evidence of interaction with an inhomogeneous medium, such as shells of CSM previously released during the LBV-phase (or disks of material, e.g. Levesque et al. 2013).

Can we associate specific features in the 2012-B light curve with known past eruptions of SN\,2009ip? If an explosion at time $t_{\rm expl}$ releases ejecta material with velocity $v_{\rm expl}$, then the expected time of interaction, $t_i$, with material from a past eruption moving at velocty $v_{\rm erupt}$ since time $t_{\rm erupt}$ is given by:

\begin{equation}
t_{i} = t_{\rm expl} +  \frac{ v_{\rm erupt} \times (t_{\rm expl}-t_{\rm erupt}) }{ v_{\rm expl} - v_{\rm erupt} }.
\end{equation}

\noindent
Smith et al. (2010) find that the bulk outflow velocity during the 2009 event was $v_{\rm erupt}$ = 550 $\rm km\ s^{-1}$. Here we consider the physical scenario of Smith et al. (2013), that a core collapse supernova occurred at the start of the 2012-A event, after which the 2012-B event was powered by the interaction of fast-moving supernova ejecta with material from past eruptions. In this scenario, the modified Julian date for the time of explosion is $t_{\rm expl}$ = 56148, and we measure $v_{\rm expl}\sim4500$ $\rm km\ s^{-1}$ from the center of the blue-side H$\alpha$ absorption in our high resolution spectrum in \S~\ref{sec:spec}. Although multiple velocity components of explosion ejecta have been observed (e.g. RM13), we use $v_{\rm expl}$ = 4500 $\rm km\ s^{-1}$ as representative of the bulk of the material. In Figure \ref{fig:lc_past}, black stars represent the magnitudes of past eruptions from Pastorello et al. (2012), plotted as a function of $t_i$, the predicted date at which eruption material would be encountered by the supernova ejecta.

Of course, we do not expect that all features will align exactly; the fastest moving ejecta will reach the CSM first, and the distribution of ejecta velocities will smear out the timescale of interactions. For example, RM13 find 3 distinct velocity components in the blue-side absorption features of H$\alpha$ from high-resolution spectra of the 2012 event, and past eruptions did not all have precisely $v_{\rm erupt}$ = 550 $\rm km\ s^{-1}$. Additionally, the material from each eruption has a unique mass and density, and the ejected material is likely to be asymmetrically distributed in the circumstellar environment (e.g. Levesque et al. 2013, Smith et al. 2013). However, in Figure \ref{fig:lc_past} the pattern of past eruptions from 2011 and 2009 does generally agree with the characteristics of the 2012-B event. The temporal spacing of the small deviations in the light curve is about 10--20 days, similar to the expected spacing of interactions between SN ejecta and past material from the 2011 eruptions. Figure \ref{fig:lc_past} also suggests that the abrupt and permanent change in the decline rate at the ``ankle'' may be related to material from the 2009 eruptions.

The opposing physical interpretation of SN\,2009ip put forth by RM13 is that the 2012-A event was a pre-cursor eruption to a stronger eruption at the start of 2012-B, which was subsequently powered by the immediate interaction material ejected in the 2012-A event. This scenario is also plausible under our current analysis. If we instead reflect the historical light curve around the start of the 2012-B event, as done in Figure \ref{fig:lc_past} for the start of the 2012-A event, we find that material ejected at the start of the 2012-B event would encounter the slower moving material from 2011 at around the time of the ``bump''. Ultimately we cannot rule out either scenario based solely on a comparison of the 2012-B light curve morphology with past eruptions. We continue this discussion of CSM interactions as a probe of past mass loss in \S~\ref{sec:spec_nHa}, where we use the H$\alpha$ narrow line emission to constrain the amount of mass lost in the 2011 eruption.

\begin{figure}[t]
\epsscale{1.2}
\plotone{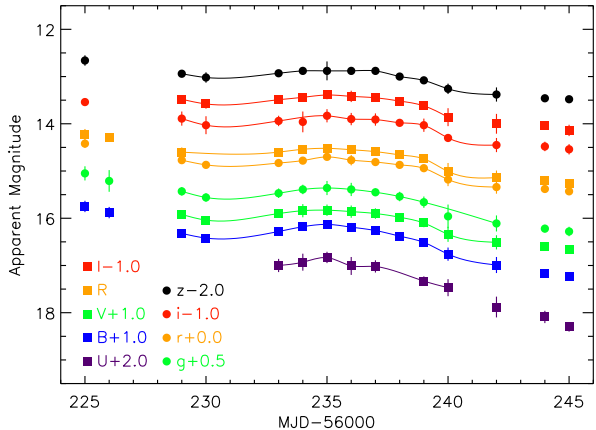}
\caption{Light curve of SN\,2009ip during the ``bump'', from FTS and CTIO in Sloan filters $gri$, $z$-short, and Johnson-Cousins $UBVRI$, with offsets added for clarity. Lines show spline fits to the photometry. All filters appear to peak on MJD=56235 (2012-11-04), with the exception of $z$-short which remains constant for the 4 days around the peak of the ``bump''.  \label{fig:lc_bump}}
\end{figure}

\subsection{A Detailed Look At The Bump}\label{sec:LC_bump}

MF13 fit splines to their {\it Swift} UV and $UBVRI$ photometry at the time of the light curve ``bump", starting on MJD $\sim$ 56232 (2012-11-01), and find that the ``bump" occurs earlier in the redder filters. The ``bump" itself is likely caused by explosion ejecta encountering an inhomogeneity in the CSM. MF13 point out that dust is not likely to cause this timing offset between UV and optical, suggesting instead that the delay is caused by the UV photons being more susceptible to scattering, and therefore taking longer to emerge. The {\it Swift} UV data published by MF13 does not exhibit a bump so much as a brief plateau before the decline resumes a couple of days after the optical peak for the bump; this may bias the ``bump" UV peak computation. While the optical and UV clearly behave differently, the difference in their $UBVRI$ filters is ambiguous, as the $UBV$ peak dates only differ by $\sim$1 day. The spline fit to $RI$ filters appears to peak several days earlier, but MF13 suggest that under-sampling in their $RI$ filters may be the true cause of this difference.

\begin{figure}[t]
\epsscale{1.2}
\plotone{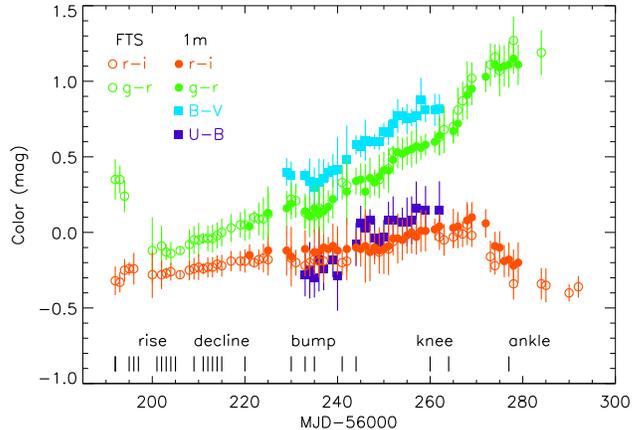}
\caption{Color evolution of SN\,2009ip during the 2012-B event in $U-B$, $B-V$, $g-r$ and $r-i$. Short black lines along the bottom note epochs of spectroscopy. \label{fig:lc_color}}
\end{figure}

With our near-daily cadence and multi-band coverage at the time of the ``bump", can we add to this discussion? In Figure \ref{fig:lc_bump} we fit our $UBVRI$ and $griz$ photometry with splines, and find the bump reaches its peak at the same day in all optical filters. In addition, Figure \ref{fig:lc_color} shows that the evolution in $B-V$ and $g-r$ colors show a brief blueward dip, coincident with the time of the ``bump'' in the optical. After this, the colors resume their steady redward progression seen over the bulk of the 2012-B event. The presence of dust would cause a bigger change in magnitude in the redder filters, as so is unlikely to contribute to the ``bump''. This ``bump'' is likely caused by the ejecta material encountering a denser region of the CSM. This is supported by the spectroscopic evolution in the H$\alpha$ line at this time: the decline rate of the flux in the narrow component flattens out at the time of the ``bump", as discussed further in Section \ref{sec:spec_Ha}.

\begin{figure}[t]
\epsscale{1.2}
\plotone{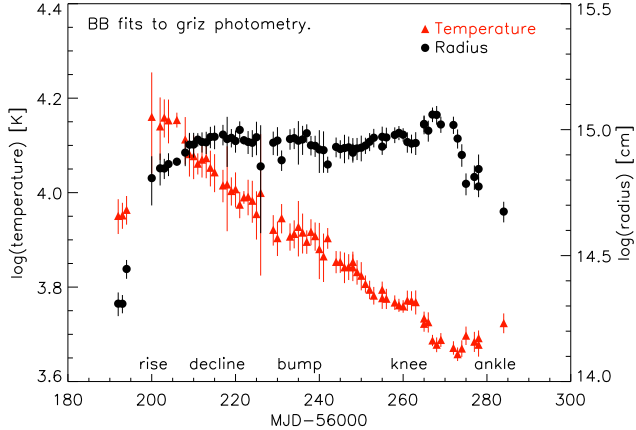}
\caption{Evolution in blackbody temperature (left axis, red triangles) and radius of the emitting region (right axis, black circles) from fits to our multi-band LCOGT photometry. \label{fig:lc_BB}}
\end{figure}

\subsection{Color Evolution and Blackbody Interpretation}\label{sec:LC_color}

In Figure \ref{fig:lc_color} we show the evolution of SN\,2009ip during the 2012-B event in four colors: $U-B$, $B-V$, $g-r$, and $r-i$. We also fit blackbody temperature and radius of the emitting region to the LCOGT $griz$ photometry, and plot the evolution in Figure \ref{fig:lc_BB}. Together, these figures show that the steady redward progression of all colors is a result of blackbody cooling. The color evolution shows several blueward deviations that coincide with bumps in the light curve. The first, at MJD$\sim$56235 (2012-11-04), coincides with the bump described in Section \ref{sec:LC_bump}. The second, at MJD$\sim$56248, is most prominent in $U-B$ and coincides with the second bump, seen most clearly in the $UBVRI$ light curve of Figure \ref{fig:lc_UBVRI}. For both events, the temperature evolution in Figure \ref{fig:lc_BB} shows a brief plateau in the cooling at these times, indicating an temporary injection of energy. In general, the color, temperature, and radius that we present agrees well with its interpretation as the explosive ejecta interacting with CSM previously released by the LBV. Material from the 2011 eruptions traveling at $v_{\rm erupt}$ = 550 $\rm km\ s^{-1}$ since MJD = 55700 would, 490 days later on MJD = 56190, be at distance $\sim$ 2.3 $10^{15}$ cm. This agrees with the radius of the blackbody emitting region having a radius $\sim10^{15}$ cm, as seen in Figure \ref{fig:lc_BB}.

After the ``knee'', starting at MJD $\sim$ 56270, the redward color evolution in $g-r$ flattens out, and in $r-i$ it takes a blueward turn. These observations agree with published color evolution for SN\,2009ip. MF13 present a flattening in all optical colors at this time, with blueward dips after MJD $\sim$ 56270 for all colors except $V-R$, for which the dip starts 5--10 days later. Similarly, RM13 show some flattening in their $B-V$ color evolution at this time, but only their UV-optical colors show such a solid bluewards turn. Unlike the blue deviations associated with small bumps in the light curve, these blueward turns are not related to an increase in temperature. Instead, at this time the light-curve decay is steeper in the $i$- and $z$-bands, as we show in Figure \ref{fig:lc_griz}. The slower decline in the $r$-band at the ``knee'' is likely influenced by the flux in the H$\alpha$ emission line which, as we present and discuss in \S~\ref{sec:spec_Ha}, is simply declining linearly at this time. RM13 remark that at this time, the near-infrared (NIR) is decaying more rapidly {\it and} there is a clear UV-excess over their black-body fits, and we agree that in general SN\,2009ip is not well fit by a black-body at this late time. For this reason, although Figure \ref{fig:lc_past} suggests that the temperature and radius evolution around the time of the ``ankle'' is related to interactions with material from the 2009 eruptions, we caution that may not be the correct interpretation.

\begin{figure}[t]
\epsscale{1.2}
\plotone{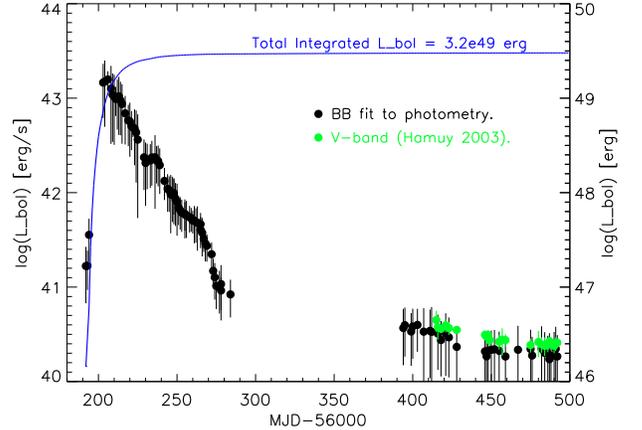}
\caption{Bolometric luminosity as a function of time for all LCOGT coverage of SN\,2009ip. Black points represent $L_{bol}$ derived from the integrated best-fit blackbody spectrum, and green points are derived from our $V$-band magnitudes and the methods of Hamuy (2003). The blue line, corresponding to the right-hand y-axis, represents the integrated bolometric luminosity output over time.  \label{fig:lc_bolo}}
\end{figure}

\subsection{Event Energetics from Bolometric Luminosity}\label{sec:LC_bolo}

We integrate our blackbody fits from \S~\ref{sec:LC_color} and plot the evolution in bolometric luminosity ($L_{bol}$) during the main 2012-B event in Figure \ref{fig:lc_bolo}. In order to understand the relative sizes of inhomogeneities in the CSM, we measure the fraction of bolometric energy in the features. To do this, we extend the rate of decline during the 10 days prior to the ``bump'' (MJD = 56220 to 56230) forward in time, and predict a $L_{bol}\sim1.3\times10^{42}$ $\rm erg \ cm^{-2} \ s^{-1}$ for MJD = 56235. On this day, the ``bump'' feature has raised the bolometric luminosity to $L_{bol}\sim2.5\times10^{42}$ $\rm erg \ cm^{-2} \ s^{-1}$ -- doubling the total energy output. This additional injection of energy, $L_{bol}\sim1.2\times10^{42}$ $\rm erg \ cm^{-2} \ s^{-1}$, is $\sim10$\% of the bolometric luminosity at peak. As discussed in \S~\ref{sec:LC_CSM}, multiple qualities of the fast-moving ejecta and the CSM together produce these bumpy features, and we cannot derive direct constraints (i.e. it does not directly follow that inhomogeneities in the CSM are 10\% of the total mass). What is clear is that the clumps in the CSM are large enough to indicate episodic eruptive mass loss occurs in the late stages of massive star evolution. Alternative explanations in include that these bolometric fluctuations are consistent with pulsations in a surviving A-type star (Martin et al. 2013), perhaps the mark of a binary system as presented in Kashi et al. (2013). Finally, we note that although these fluctuations appear to be unique for SN\,2009ip, the relatively sparse data sampling in past events such as PTF10tel, SN\,1998S, and SN\,1999el shown in Figure \ref{fig:lc_comp} could hide such features. 

In Figure \ref{fig:lc_bolo} we extend the bolometric light curve out to late times. Since the photometry is not very well fit by a blackbody at late times, we also use the method of converting a core collapse supernova's late-time $V$-band magnitude to bolometric luminosity presented in Hamuy (2003). The results are plotted as green points in Figure \ref{fig:lc_bolo}, and agree within the (rather large) uncertainties of $L_{bol}$ derived from blackbody fits. In addition, we show how the integrated bolometric luminosity builds up over time (blue line), with most of the energy released within the first $\sim$20 days, and a final total energy output of $\sim$3.2$\times$$10^{49}$ ergs.

Under the simplifying assumption that the bulk of the emission originates in the kinetic energy of the ejecta with a bulk velocity of $v_{expl}$ = $\sim$4500 $\rm km\ s^{-1}$, the mass ejected in the 2012 explosion is $\sim0.1$ $\rm M_{\odot}$. This is only a small fraction of the supposedly 50--80 $\rm M_{\odot}$ progenitor star \citep{Smith2010}, and suggests that an insufficient amount of the star was unbound in the explosion for it to be a true supernova. However, NS13 point out that this is actually a minimum ejecta mass estimate because the CSM is not isotropic (e.g. Levesque et al. 2013), allowing much of the ejecta to escape, and the main brightening may be caused by only the fastest-moving ejecta. NS13 furthermore point out that 0.1 $\rm M_{\odot}$ of ejecta would be optically thin within days, and that this is directly contradicted by the broad H$\alpha$ emission component that is present until late in the 2012-B event, as we show in Figure \ref{fig:halpha}. They argue that this indicates ejecta masses of at least 4--6 $\rm M_{\odot}$.

\begin{figure}[t]
\epsscale{1.2}
\plotone{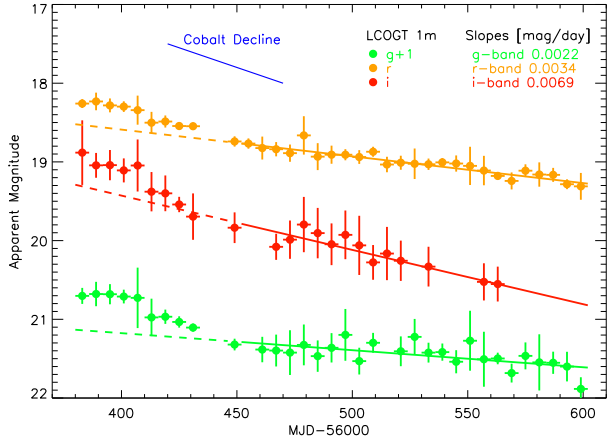}
\caption{The late time decline of SN\,2009ip in 6 day bins for SDSS filters $gri$ (colored circles). Solid lines represent the best linear fit to data later than MJD=56450; dashed lines extrapolate the fit backward in time to show the non-linear decline in all bands. The slope of $^{56}$Co decay is shown in blue for reference. \label{fig:lc_late}}
\end{figure}

\subsection{Power Source of the Late-Time Decline}\label{sec:LC_late}

When SN\,2009ip emerged from behind the sun in May 2013, we resumed our photometric monitoring program in $BVgri$ using the LCOGT 1.0m network. The calibrated photometry for $gri$ is shown in Figure \ref{fig:lc_late}. Although SN\,2009ip is steadily declining, the slope is not constant: on MJD $\sim$ 56410, a steeper decay event began and lasted $\sim$20 days. This event is steeper in the $g$- and $i$-bands compared to the $r$-band filter, as was the ``knee" in Figure \ref{fig:lc_griz}. To emphasize this dip, in Figure \ref{fig:lc_late} we make linear fits to the photometry for dates after MJD = 56440 (solid lines) and extend them backwards (dashed lines). The best fit slopes in all bands after this dip are $<$0.005 magnitudes per day, flatter than the typical decay of $^{56}$Co, which is $\sim0.01$ magnitudes per day. The combined evidence of features in the late-time light curve {\it and} a slope shallower than a $^{56}$Co decay indicates the optical emission from SN\,2009ip is still influenced by ongoing CSM interaction. This is not uncommon for SN\,IIn, as shown in Figure \ref{fig:lc_comp}, and contributions from $^{56}$Co decay from a supernova cannot be excluded. As a test, if we include only dates 56390 $<$ MJD $<$ 56430 in the linear fit, the slopes increase to 0.01, 0.009, and 0.019 magnitudes per day in $g$, $r$, and $i$ bands respectively. With less complete late-time coverage a late-time slope consistent with $^{56}$Co decay could be derived. Without LCOGT's high cadence observations at late times, we may well have drawn different conclusions about the late-time decline of SN\,2009ip.

When the late-time decline of supernovae is powered by the decay of $^{56}$Co to $^{56}$Fe, which has an e-folding time of 111 days (or a 77 day half-life), the decline rate is $\sim$0.01 magnitudes per day. Type IIn SNe are often seen to remain bright and decline slower than this at late times for a long time (e.g. Zhang et al. 2012), due to the heated CSM. This is exemplified by SN\,1988Z and SN\,1995G in Figure \ref{fig:lc_comp}, and also by SN\,2009ip in Figure \ref{fig:lc_late}. The only place in the SN\,2009ip light curve where $^{56}$Co decay could have been the dominant source of optical emission is the $\sim$100 day window between MJD = 56290 and MJD = 56390 when SN\,2009ip was behind the sun. During this time, the $r$-band magnitude declined from 17.5 to 18.6 magnitudes, which is an average of 0.01 magnitudes per day. We can only use this decline to estimate an upper limit on the potential mass of $^{56}$Ni synthesized in the explosion. To do so, we use our bolometric luminosity, adopt the earliest possible explosion date (the start of the 2012-A event on MJD = 56150), and use the expression for nickel mass from Hamuy (2003). We find $M_{Ni}<0.04$ $\rm M_{\odot}$, consistent with previously reported estimates (e.g. MF13). However, for SN\,2009ip this estimate of potential $^{56}$Ni mass is not very restrictive on the explosion mechanism. 

Future monitoring will reveal whether the decline of SN\,2009ip exhibits further features of interaction, or continues to steadily decline. The faintest recorded historical magnitude since detection as a transient is $m_V=21.46$ magnitudes on 2009-11-24 (AP13), and in archival {\it HST} images it was measured to be $m_{F606W}=21.8$ magnitudes \citep{Smith2010}. At the current rate of $r$-band decline, SN\,2009ip will not pass this magnitude until mid-January 2016. A complete disappearance would confirm that these events were powered by a terminal supernova. Given that massive stars are so frequently in binary systems with other massive stars, this may not happen \citep{Sana2012}. If half of the flux in the archival {\it HST} images actually came from a binary companion, the future magnitude of the remaining companion may be $m_{F606W}=22.6$, easily detectable with {\it HST}.

\begin{figure}[t]
\epsscale{1.2}
\plotone{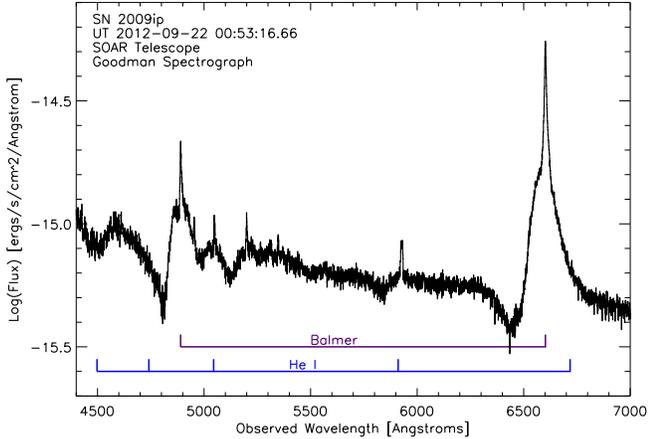}
\caption{Spectra of SN\,2009ip on 2012-09-22, before the rise of the 2012-B event, from the Goodman spectrograph at SOAR Telescope. We identify lines of the hydrogen Balmer series (purple), and singly-ionized helium (blue). \label{fig:SOAR_spec}}
\end{figure}

\begin{figure*}[t]
\epsscale{1.17}
\plottwo{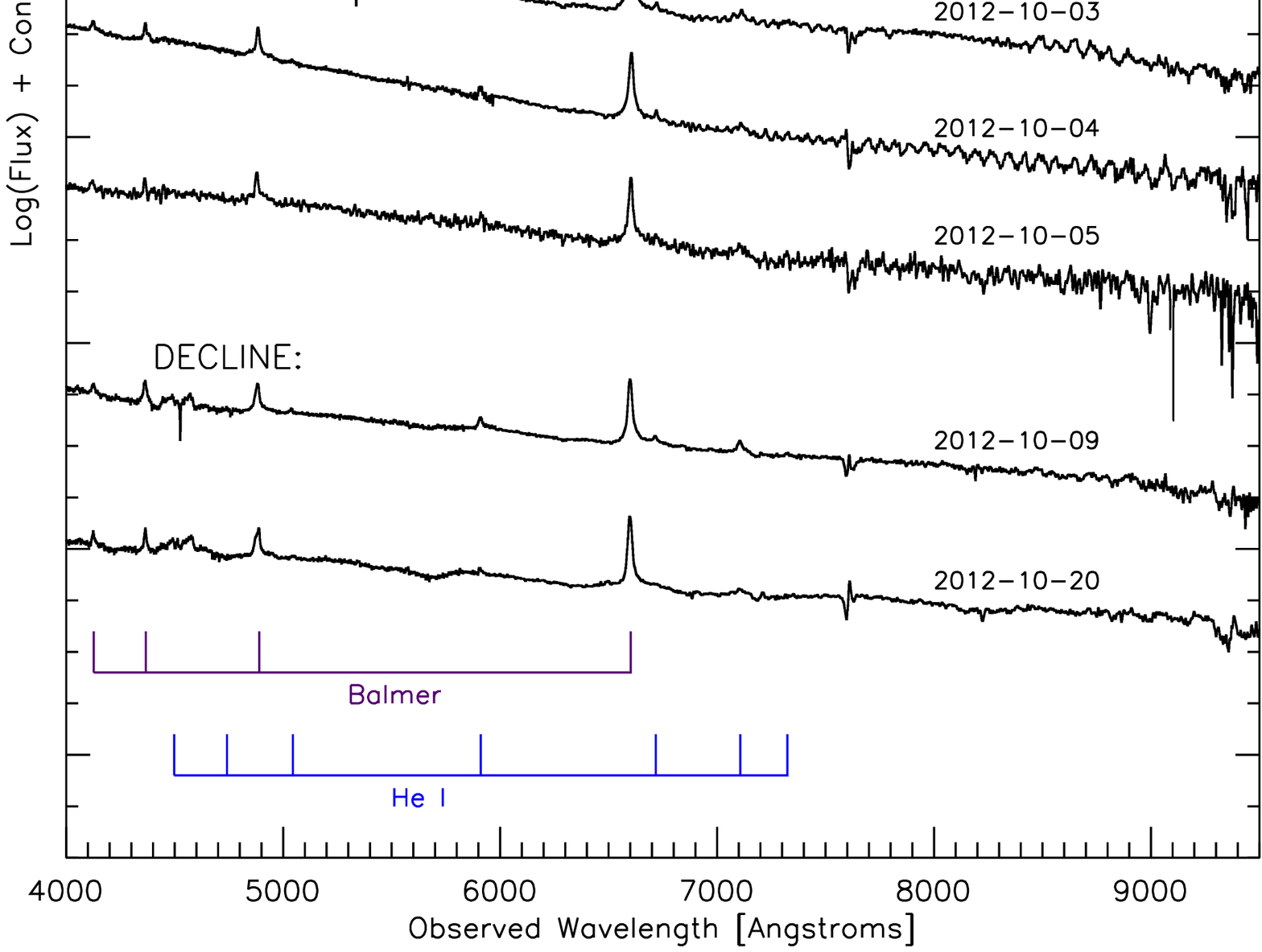}{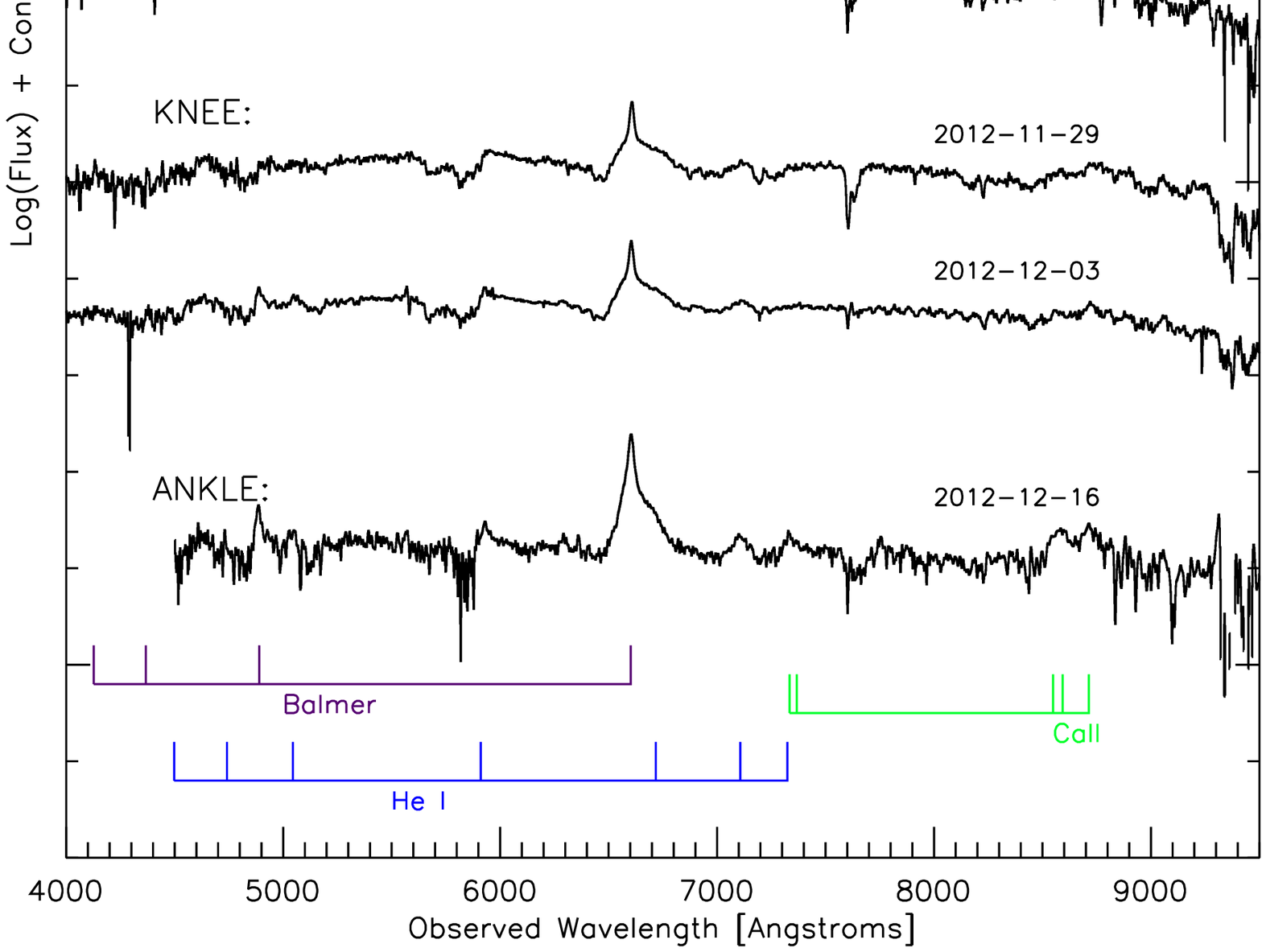}
\caption{\textbf{Left:} Time series of FLOYDS spectra for SN2009ip during the rise and decline eras of the light curve, plotted in $\log(flux) + C$ for clarity. Positions of emission lines (at the host galaxy's redshift) from Hydrogen Balmer series, singly ionized Helium, and doubly ionized Calcium are labeled in purple, blue, and green respectively. The Earth symbol denotes regions of increased noise where atmospheric telluric lines have been removed. \textbf{Right:} As for the left plot, but for the bump, knee, and ankle eras of the light curve. \label{fig:FLOYDS}}
\end{figure*}

\begin{figure}[t]
\epsscale{1.15}
\plotone{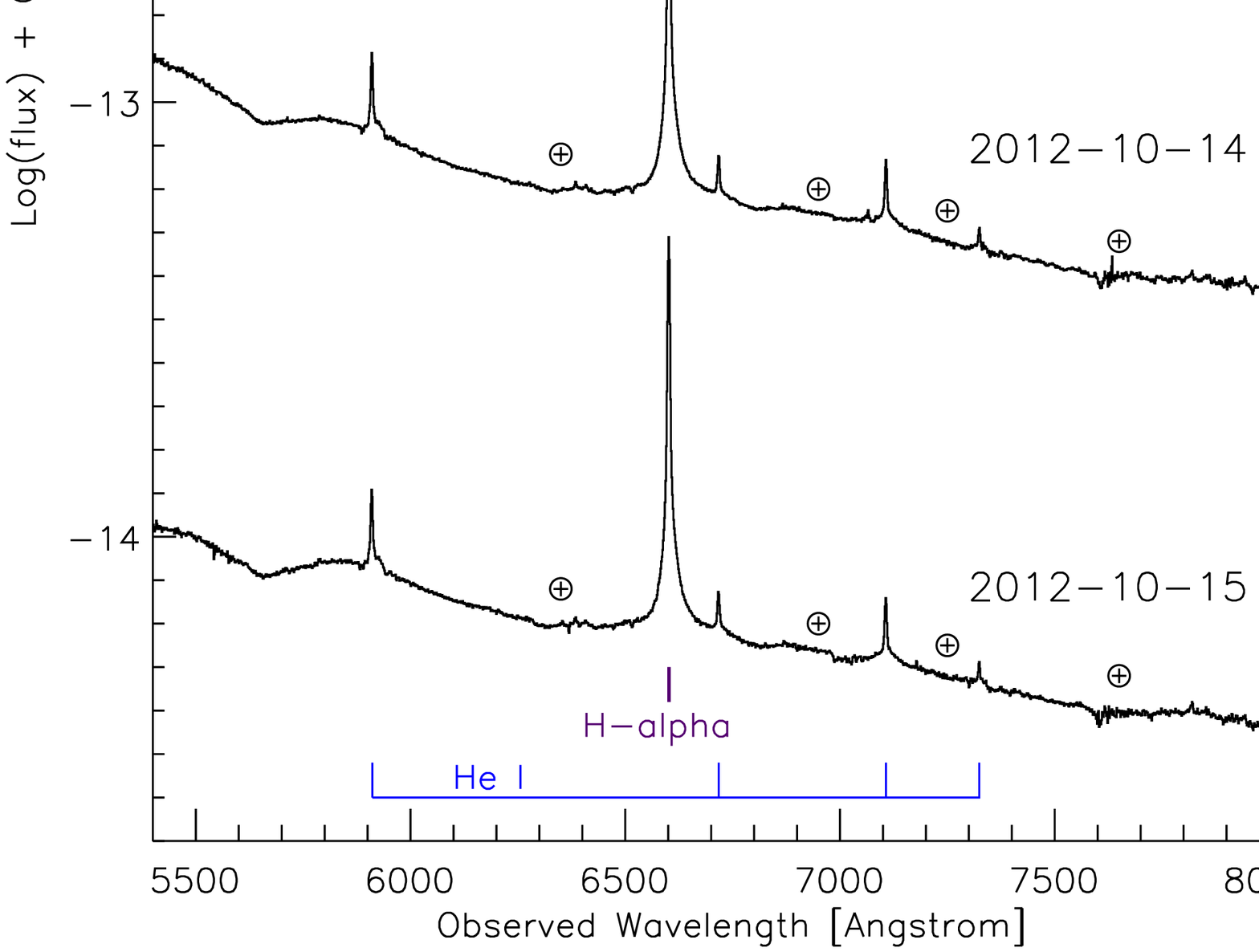}
\caption{Spectra of SN\,2009ip from the KPNO Mayall 4m Telescope. The magnitude of SN\,2009ip was effectively constant during this time, and the spectra are identical -- without arbitrary offsets they cannot be distinguished from each other. These three spectra fit in between the two ``decline" phase spectra in Figure \ref{fig:FLOYDS}. The Earth symbol denotes regions of increased noise where atmospheric telluric lines have been removed.  \label{fig:kpnohr}}
\end{figure}

\section{Time Series Spectroscopy of SN\,2009ip}\label{sec:spec}

Our spectroscopic time series for SN\,2009ip begins on 2012-09-22, at the start of the 2012-B event, when we obtained a spectrum with the Goodman High Throughput Spectrograph at SOAR Telescope, as shown in Figure \ref{fig:SOAR_spec}. At this point, the Balmer lines clearly show narrow and broad components and blue-shifted absorption. The narrow component is a Lorentz profile with FWHM $\sim10$ \AA\ and a peak wavelength of 6600 \AA. The broad component is a Gaussian profile with FWHM $\sim130$ \AA\ and a peak wavelength of 6590 \AA, and the absorption component is fit by a single Gaussian with a FWHM $\sim150$ \AA\ and a central wavelength of 6500 \AA. The absorption component extends to $\sim6300$ \AA, evidence of fast moving material at velocities $\sim$13500 $\rm km\ s^{-1}$, but the bulk of the material is at velocity $\sim$4500 $\rm km\ s^{-1}$. Aside from the Balmer lines, the second most prominent feature is the singly ionized Helium lines, indicating the presence of He I in the circumstellar material, which is unsurprising for a massive star progenitor.

As SN\,2009ip rose and fell, we continued to monitor it with the FLOYDS spectrograph. These spectra are shown in Figure \ref{fig:FLOYDS}, and listed in Table \ref{tab:spec}. In the week after peak light we obtained a series of spectra 5 nights in a row with the RC spectrograph on the Mayall telescope at KPNO; the 3 taken in the higher resolution setup are shown in Figure \ref{fig:kpnohr}. The changing slope of the continuum represents a cooling blackbody, the result of CSM interaction as discussed for the photometry in Section \ref{sec:LC_color}. During the spectral evolution of SN\,2009ip, the most conspicuous feature is H$\alpha$, which shows a changing combination of broad and narrow emission, and sometimes multiple blue-side absorption features. The rest of the hydrogen Balmer series, and narrow emission lines from singly-ionized helium, persist throughout the FLOYDS data set. These emission lines are generated by the circumstellar hydrogen and helium from previous mass-loss events during the LBV-phase of the massive progenitor star, with which the explosion ejecta is colliding.

By the end of our FLOYDS spectroscopic time series (2012-12-16), SN\,2009ip had returned to nearly the same brightness as when we started (2012-09-25), as shown in Figure \ref{fig:lc_griz}. After the ``ankle'', the brightness of SN\,2009ip appeared to abruptly stop declining. This photometric change could be interpreted as an LBV star returning to quiescence after a non-terminal outburst, but the spectrum reveals otherwise: it is no longer dominated by narrow lines, but has not returned to its pre-outburst appearance. This is most clear in the appearance of forbidden calcium emission in the ``ankle" spectrum of Figure \ref{fig:FLOYDS}. Shortly thereafter, SN\,2009ip disappeared behind the sun. Upon its return we obtained the late-time spectra with Gemini Observatory, and found that the forbidden calcium lines had grown in strength.

In the following sections, we use our time series of spectra to constrain the physical explosion scenario for SN\,2009ip. We use the H$\alpha$ line luminosity to place an upper limit on the CSM mass involved in interactions with the fast-moving ejecta in \S~\ref{sec:spec_nHa}; we analyze the evolution of the H$\alpha$ profile during the 2012-B event and examine the relative contributions from its narrow, broad, and absorption components in \S~\ref{sec:spec_Ha}; and we discuss this late-time spectrum in context with spectra for other energetic events in \S~\ref{sec:spec_late}.

\subsection{Constraints on Past Mass Loss from Narrow H$\alpha$}\label{sec:spec_nHa}

Under the assumption that the 2012-B event is powered by fast-moving ejecta interacting with a shell of material from the 2011 eruption, as discussed in \S~\ref{sec:LC_CSM}, we can constrain the amount of mass lost during that episode. Ofek et al. (2013b) provide a thorough examination of multi-wavelength mass loss rate indicators for SN\,2009ip. For our calculation, we follow their Equation 4 for the mass of hydrogen, $M_H$, from the luminosity of the narrow component of H$\alpha$, $L_{H\alpha}$:

\begin{equation}
M_H = \frac{m_p \ L_{H\alpha}}{h \ v_H \ \alpha^{eff}_H \ n},
\end{equation}

\noindent
where $m_p$ is the proton mass, $h$ is Planck's constant, and $\alpha^{eff}_H$ is the recombination coefficient at 10000 Kelvin, which is appropriate for our use as we find a blackbody temperature of $\sim10000$ K in \S~\ref{sec:LC_color}. The term $n$ is the particle density profile, which we express as:

\begin{equation}
n = \frac{M_H/m_p}{V},
\end{equation}

\noindent
where the volume $V$ is the shell occupied on MJD = 56192 by material ejected at 550 $\rm km\ s^{-1}$ for the 200 days starting on MJD = 55700, which was the duration of the 2011 eruptions. At the peak of the 2012-B event, the flux in the narrow component of H$\alpha$ was $F_{H\alpha}\sim5\times10^{-13}$ $\rm erg\ cm^{-2} \ s^{-1}$, which we convert to a luminosity using a distance of 25.2 Mpc as in \S~\ref{sec:LC}. We find that $M_H$ $\sim$ 0.1 $\rm M_{\odot}$ of material was released during the 2011 eruptions, but note that this is an upper limit, as it relies on the assumption that all of the hydrogen was ionized by the 2012-B ejecta. The implied mass loss rate during the 2011 eruptions is $\sim$0.18 $\rm M_{\odot}\ yr^{-1}$. As described in Ofek et al. (2013b), this upper limit is significantly higher than mass loss rates derived from e.g. X-ray observations, but this discrepancy is important because it supports an asymmetric distribution of CSM (e.g. a disk, as in Levesque et al. 2013). For example, if the CSM is in a disk with $\sim$10\% the volume of the assumed shell, then $M_H$ $\sim$ 0.01 $\rm M_{\odot}$.

\begin{figure}[t]
\epsscale{1.0}
\plotone{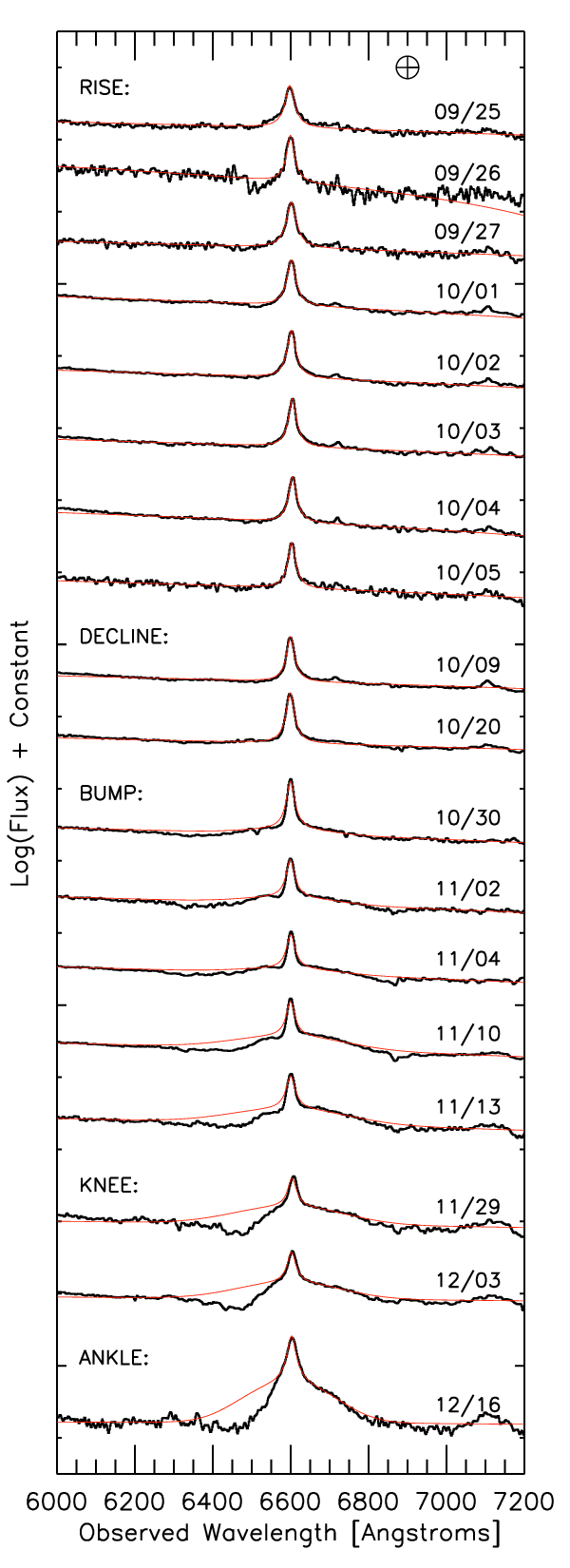}
\caption{Evolution of the H$\alpha$ line from FLOYDS time-series spectra, labeled with date and phase era (rise, decline, bump, knee, and ankle). Red lines represents fits to the data: from 09/25 to 10/20 a single narrow Lorentz profile is fit, after which a broad Gaussian component is also fit (to the red side only). The blue side shows increasingly significant multi-component absorption features, which we have purposely omitted from the red line fit in order to give a better visual impression of their impact on the broad emission component. The Earth symbol denotes wavelengths of increased noise where atmospheric telluric lines have been removed.  \label{fig:halpha}}
\end{figure}

\subsection{The Physical Source of H$\alpha$ Evolution}\label{sec:spec_Ha}

The H$\alpha$ line exhibits three main components in varying strengths: narrow emission, broad emission, and blue-side absorption. These components represent the physical evolution of the ejecta and CSM interactions powering the 2012-B event of SN\,2009ip. In Figure \ref{fig:halpha}, we show a time series of the H$\alpha$ line from our LCOGT FLOYDS spectroscopic series, labeled with the date and the phase era (rise, decline, bump, knee, and ankle). The five spectra taken with the KPNO Mayall RC spectrograph between 10/11--10/15 fit into the ``decline'' era after 10/09. In the KPNO spectra, the H$\alpha$ emission is dominated by a narrow Lorentzian profile from the CSM emission, and Figure \ref{fig:kpnohr} shows how there is effectively no evolution in the spectrum, just as the photometry is nearly constant at this time in Figure \ref{fig:lc_griz}.

To the H$\alpha$ emission we fit a single narrow Lorentz profile for dates up to 10/20, after which this is combined with a broad Gaussian profile (after first subtracting the continuum). Blue-shifted absorption is clearly present on 09/26 and after 10/30, but exhibits multiple components so we do not fit the blue-shifted absorption. The broad Gaussian emission is fit to the red side only, with a peak wavelength restricted to within 5 \AA\ of the narrow line peak. In Figure \ref{fig:halpha}, the red line represents this fit to the narrow and broad emission only, because we find that not including the absorption components has actually highlighted its shape and contribution to H$\alpha$. In the following two sections, we discuss the possible physical sources of the emission and absorption components in turn.

\subsubsection{Narrow and Broad Emission Features}

Over the 2012-B event, the width of the narrow component of H$\alpha$ appears to increase from FWHM $\sim$ 700 to 1000 $\rm km\ s^{-1}$, or $\sim$ 16 to 20 \AA, but this is within uncertainty. As the continuum and the flux in the narrow component decline, the broad component re-emerges at the start of the ``bump" era. The width of the best fit broad component appears to increase from $\sim$ 10600 $\rm km\ s^{-1}$ on 2012-10-30 to $\sim$ 15000 $\rm km\ s^{-1}$ on 2012-11-13, encompassing the duration of the ``bump'', after which it appears to decreases back to $\sim$ 10600 $\rm km\ s^{-1}$. 

\begin{figure}[t]
\epsscale{1.2}
\plotone{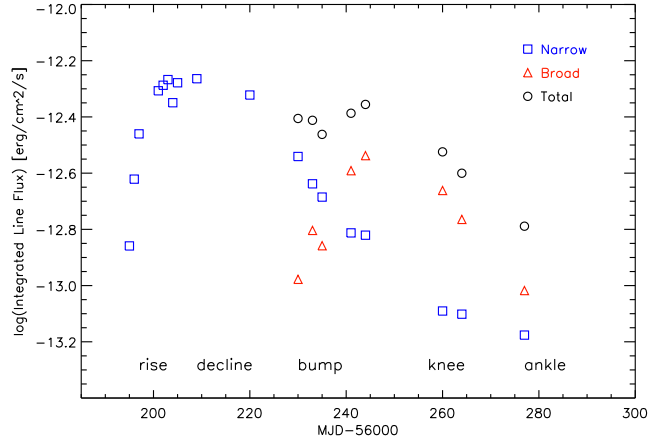}
\caption{Total integrated flux in the H$\alpha$ emission line components: the narrow Lorentz profile (blue squares), the broad Gaussian profile (red triangles), and the sum total when both components contribute (black circles). Eras of the light curve features, defined in Figure \ref{fig:lc_griz}, are written across the bottom.   \label{fig:halpha_flux}}
\end{figure}

The evolution of the integrated, continuum-subtracted flux in the fit narrow and broad components over the course of the 2012-B event are plotted in Figure \ref{fig:halpha_flux}. The flux in the narrow component rises and declines similar to the light curve -- as it should, because the continuum emission is also dominated by CSM interaction. The decline of the flux in the narrow component exhibits a plateau at the time of the ``bump'', which is consistent with our interpretation of the ``bump" as explosion ejecta interacting with an inhomogeneity in the CSM. Line emission as broad as that seen at post-``bump" eras of SN\,2009ip is typically associated with fast expansion, and this leaves us with several questions: what is the physical cause of the broad emission? Why does its width and integrated flux appear to increase and decrease? How is it related to the light curve ``bump"?

The coincidence of the ``bump" with an increase in the relative contribution from the broad component seems difficult to resolve with the underlying cause of the ``bump" being the result of ejecta encountering an inhomogeneity in the CSM, which is the source of narrow emission. MF13 point out that interaction with a clump of CSM can cause broader emission through photon scattering. If so, we might expect to see a similar increase in the relative contribution from the broad component at the time of the ``knee" and/or ``ankle" -- but Figure \ref{fig:halpha_flux} has only a few data points at those times, and they suggest a continuing decline. Given that the ``knee" is a lower energy feature, and that our spectroscopic sampling at this phase is insufficient to see short time-scale changes in the H$\alpha$ broad component, we cannot confirm the photon scattering hypothesis.

NS13 interpret the 2012-A event as a true supernova, and the source of the broad component of H$\alpha$ seen at the start of 2012-A and during the decline of 2012-B. With this model the broad component always there, but during the 2012-B event is swamped with continuum and narrow line emission from the subsequent interaction between supernova ejecta and CSM. As the continuum luminosity declines, the broad component from the SN is again revealed, and it is simply a coincidence that it re-emerges at the same time as the ejecta encounters a CSM clump and creates the ``bump". In this case, the width and luminosity of the broad component may only appear to rise and decline because the continuum is receding. NS13 further support this interpretation by showing the H$\alpha$ equivalent width, $EW = | \int 1-(F_{\lambda}/F_0) d\lambda |$, is constantly increasing, as is commonly seen for supernovae (NS13). We also measure the equivalent width of H$\alpha$ for FLOYDS spectra, and find it increases from 100 to 1000 \AA\  from the ``rise'' to the ``ankle'', in agreement with NS13. Note that a continually increasing $EW$ is not at odds with the continually declining integrated line flux in Figure \ref{fig:halpha_flux}: this happens in the case where the continuum flux is declining faster than the line flux.

There remains the third interpretation that the broad emission is generated by another non-terminal eruption of the LBV star. This would explain both the apparent rise and decline of the broad component emission, and the appearance of multiple new velocity components of H$\alpha$ absorption presented by RM13 (discussed further in \S~\ref{sec:absfeat}). In the scenario of continuing eruptions of an LBV-phase star, this would be third in a line of three eruptions spaced $\sim$40 days apart. Although the photometric evolution of the 2012-B event is quite distinct from the LBV-phase, the 40-day spacing is similar to that seen during the 2011 eruptions in Figure \ref{fig:pastorello}. In this scenario, the light curve ``bump" would not represent a CSM inhomogeneity, but newly ejected material interacting with the CSM. However, we are skeptical that the widths implied by our fits to the broad component are $\sim$twice that shown in the first spectrum from 2012-09-22 in Figure \ref{fig:SOAR_spec}. How could there be a third powerful eruption without a correspondingly large increase in the continuum luminosity? We suspect that the simplest explanation is that the fit width and integrated flux of the broad component is compromised by the receding continuum during this era.

\subsubsection{Absorption Features of the Ejecta Components}\label{sec:absfeat}

In Figure \ref{fig:halpha}, the mismatch between data and fit on the blue side emphasizes the complex shape of the H$\alpha$ absorption profiles. After the ``bump", the blue absorption feature extends to at most $\sim$6300\ \AA, or $\sim$15000 $\rm km\ s^{-1}$, just as it did in the 2012-09-22 spectrum of Figure \ref{fig:SOAR_spec}. Such high velocity ejecta is typical of supernova explosions, but not commonly seen for LBV -- except in a spectrum of SN\,2009ip from 2011 presented in AP13, so it is also not totally unprecedented in non-terminal events. There is also a low velocity absorption component at $\sim$ 2000 $\rm km\ s^{-1}$, most clearly seen starting on 11/02.

The appearance of lower velocity components as time progresses is commonly interpreted as due to the photosphere receding into deeper (slower) levels of the explosion. Under the interpretation of NS13, this absorbing material and its P-Cygni profile has been present continuously, but was swamped by CSM emission from peak light to around the time of the ``bump". RM13 examine high resolution spectra of the blue-shifted absorption features and find they represent three distinct velocities ($\sim$12000 km/s, $\sim$5500 km/s, and $\sim$2500 km/s), not a continuum, as would be expected for photospheric recession. These three components appear successively at 9, 28, and 60 days after peak brightness, the latter two roughly corresponding to the ``bump'' and ``knee''. This may be consistent with the interpretation of Martin et al. (2013) of the light curve features as continued pulsations of the (undead) star. Alternatively, regardless of the explosion type of the primary star, these absorption components may be jets of material released by the binary companion star \citep{Tsebrenko13}.

\begin{figure}[t]
\epsscale{1.2}
\plotone{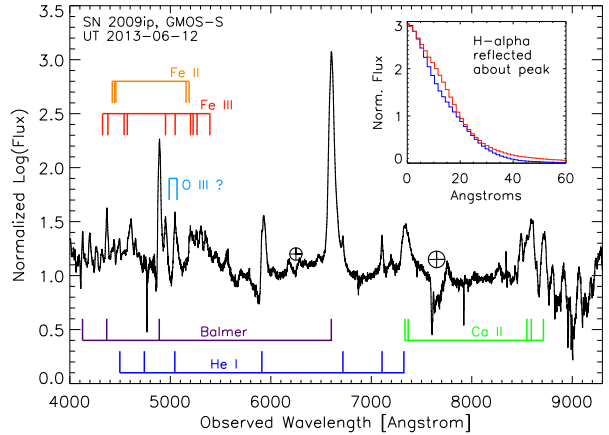}
\caption{Late-time spectrum of SN2009ip, presented in normalized $\rm \log{(flux)}$. We mark the following spectral lines: hydrogen Balmer series (purple); singly-ionized helium (blue); doubly-ionized calcium (green); and doubly- and triple-ionized iron (orange and red). We also mark the position of triply-ionized oxygen (light blue), which coincides with Fe III, and was identified in the late-time spectra of SN\,1998S \citep{Pozzo04}. The inset plot shows the continuum-subtracted H$\alpha$ line flux, reflected about the peak wavelength of the emission, with blue and red lines representing the blue and red sides. At these late times the H-$\alpha$ emission is slightly asymmetric, showing extra flux on the red side. The Earth symbols mark the positions of unremoved telluric lines. \label{fig:spec_late}}
\end{figure}

\begin{figure}[t]
\includegraphics[trim=0.5cm 0.5cm 0.1cm 1.4cm,clip=true,width=9cm]{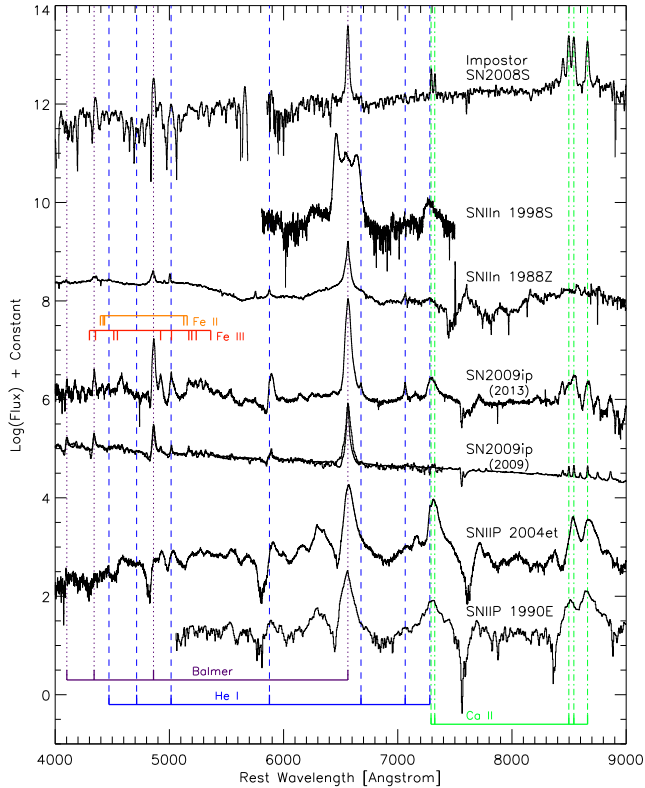}
\caption{Late-time spectrum of SN\,2009ip at 249 days since the peak of the 2012-B event, compared with late time spectra of two SNe\,IIn, two SNe\,IIP, an LBV-phase spectrum of SN\,2009ip, and a SN impostor. From top to bottom:
SN Impostor 2008S at $\sim$223 days (Botticella et al. 2009; Smith et al. 2009),
SN\,IIn 1998S at  $\sim$300 days \citep{Pozzo04}, 
SN\,IIn 1988Z at $\sim$115 days \citep{Turatto1993},
SN\,2009ip at 249 days (this paper),
SN\,2009ip from 2009 presented in AP13 (two spectra described further in text),
SN\,IIP 2004et at $\sim$250 days \citep{Sahu2006},
and SN\,IIP 1990E at $\sim$250 days \citep{Gomez2000}.
All spectra have been normalized, converted to the rest frame of the host, and offsets were added for clarity. We mark the following spectral lines: hydrogen Balmer series (purple dotted line); singly-ionized helium (blue dashed line); doubly-ionized calcium (green dot-dash line); and doubly- and triple-ionized iron (orange and red). \label{fig:spec_late_comp}}
\end{figure}

\subsection{The Late-Time Spectrum}\label{sec:spec_late}

We obtained a late-time spectrum of SN\,2009ip, shown in Figure \ref{fig:spec_late}, with the Gemini Multi-Object Spectrograph on Gemini South Telescope on 2013-06-12, when SN\,2009ip was 249 days past the maximum brightness of the 2012-B event. At this time it is similar to the spectrum taken on 2012-12-16 (the ``ankle) in Figure \ref{fig:FLOYDS}, except that at these later times and in a higher resolution spectrum, the contributions from Fe and O are more obvious, and the He and individual Ca lines more distinguished. In Figure \ref{fig:spec_late_comp} we compare the late-time spectrum of SN\,2009ip with spectra from a SN impostor, SNe\,IIn, and SNe\,IIP taken at similarly late phases, and also with LBV-phase spectra of SN\,2009ip. These spectra were obtained from the Online Supernova Spectrum Archive (SUSPECT\footnote{www.suspect.nhn.ou.edu/$\sim$suspect}) and the Weizmann Interactive Supernova Data Repository (WISEREP\footnote{www.weizmann.ac.il/astrophysics/wiserep}; Yaron \& Gal-Yam 2012). 

The late-phase spectrum of SN\,2009ip is similar to SN\,2008S, which was defined as a SN Impostor by Smith et al. (2009). SN\,2009ip has a flatter continuum flux and stronger helium lines, but the H$\alpha$ and calcium lines are equivalently strong. The major difference between SN\,2009ip and SN impostor 2008S is the presence of a broad H$\alpha$ component for SN\,2009ip, indicating the high expansion velocities of a true SN, as discussed in NS13. Such broad emission is not seen for SN Impostors.

At phases of $\sim$100 days, MF13 find the spectrum of SN\,2009ip is well matched to SN\,1998S, a peculiar SN\,IIn (e.g. Fassia et al. 2000). In Figure \ref{fig:spec_late_comp} we compare with the $\sim$300 day spectrum of SN\,1998S \citep{Pozzo04}. SN\,1998S shows a strong, broad Ca II emission, similar to SN\,2009ip. The triple peak in the H$\alpha$ line had emerged by day $\sim230$, as shown by Gerardy et al. (2000), who suggest the triple velocity components are caused by a ring-like disk structure in the CSM. This type of disk has also been suggested for SN\,2009ip based on the Balmer decrement during the 2012-B event \citep{Levesque2013}. The fact that we do not see obvious multiple velocity components does not mean there is no disk, as orientation and relative velocities may differ from SN\,1998S. In the inset of Figure \ref{fig:spec_late} we look for asymmetry in the H$\alpha$ line by reflecting it about the peak wavelength and comparing the blue and red sides. The excess on the red side could be attributed to asymmetry in the H$\alpha$ emitting region, or as a depression in the blue side due to dust formation. Future confirmation of the CSM geometry will provide better constraints on the energy budget and the true explosion mechanism.

In Figure \ref{fig:spec_late_comp}, we also compare our late-time Gemini spectrum with spectra for a typical SNe\,IIn,
SN\,1988Z at 115 days post-max \citep{Turatto1993}. At 1--4 months past peak this object has similar spectral features to those shown by SN\,2009ip in Figure \ref{fig:FLOYDS}, including asymmetry and blue-shifted absorption in the H$\alpha$ feature. At these late times, their spectra are quite similar, but SN\,2009ip shows stronger, broader lines of He I, and stronger and broader lines of Fe III and Ca II. At the chosen phase, SN\,IIn 1988Z still retains a broad component similar to that exhibited by SN\,2009ip at earlier epochs.

For comparison with the next most similar kind of core collapse supernovae, we also show two late-time spectra from SNe\,IIP: SN\,2004et at $\sim$250 days \citep{Sahu2006} and SN\,1990E at $\sim$250 days \citep{Gomez2000}. Although similar to SN\,2009ip in the presence of hydrogen and Ca II, the SNe\,IIP show broader Balmer lines with P-Cygni profiles, and show stronger Ca II. In addition they do not show the prominent He I, and have a weaker continuum in the blue end -- a combination of a cool blackbody spectrum and possibly dust production.

We also show a comparison to two spectra taken during the 2009 LBV-phase events of SN\,2009ip: the thick line is 2009-09-25 and the thin line is 2009-10-22, both from Figure 5 of AP13. On 09/25, SN\,2009ip was just declining from its second detected outburst of the year, at $R\sim19$ magnitudes. By 10/22, it had been in a plateau at $R\sim20$ for 20 days, and the H$\alpha$ emission had broadened slightly and lost its blue-shifted absorption feature. It is clear that SN\,2009ip has not yet returned to its old quiescent self, as the Fe III, He I, and Ca II emission lines are all currently broader and stronger, and more similar to SN Impostors and late-phase SN\,IIn. Only time will tell if SN\,2009ip returns to it's pre-2012, LBV-phase spectral signature.

\section{Conclusion} \label{sec:con}

In this work we have used high cadence LCOGT data, with supporting observations from other facilities, to draw the following four main conclusions about SN\,2009ip. 
(1) From our daily photometric monitoring, we observe photometric fluctuations from interactions between fast-moving ejecta and the CSM that reveal inhomogeneities in the CSM distribution. The light curve of the 2012-B event appears congruous with the timescales of past eruptions in 2011 and 2009.
(2) The peak brightness and photometric evolution of SN\,2009ip is more similar to SNe\,IIn than LBVs or SN Impostors. Our long-term monitoring shows SN\,2009ip continues to decline slowly with small fluctuations in the slope, which is shallower than if powered by the decay of Co$^{56}$. This is similar to both SNe\,IIn and SN Impostors, but SN\,2009ip has not exhibited any more LBV-like outbursts. (3) Our low and high resolution spectra during the 2012-B event exhibit evolving H$\alpha$ emission and absorption components. As the continuum and narrow line flux from CSM interaction fades, the broad emission component persists. This is characteristic of SN-like expansion velocities, but not SN Impostors or LBVs.
(4) The late-time spectrum of SN\,2009ip exhibits forbidden calcium lines which are more similar to SNe\,IIP, SNe\,IIn, and SN Impostors than the LBV-phase spectrum of SN\,2009ip. Asymmetry in the H$\alpha$ line at late times hints at dust formation and/or asymmetry in the CSM, such as a disk.

SN\,2009ip clearly fits into the class of SNe\,IIn; if the progenitor has not suffered a terminal explosion, then perhaps not all SNe\,IIn are actual supernovae. Multiple papers have remarked on the striking similarity -- right down to the timing and magnitude of a precursor outburst -- between SN\,2009ip and the SN\,IIn 2010mc (e.g. RM13, NS13). Far from a serendipitous match, Ofek et al. (2014) make a careful analysis of the Palomar Transient Factory data and find that such precursor events are common for SNe\,IIn. SN\,2009ip is unique in that it was observed first as a variable star undergoing eruptive mass loss for years before becoming a supernova. The resulting interactions of explosion ejecta with the eruption material reveal that the late stages of mass loss in massive stars can produce a CSM that has large inhomogeneities and a potentially asymmetric distribution.

SN\,2009ip is a unique and interesting event for which multiple groups have assembled a high cadence photometric and spectroscopic coverage from a wide variety of telescopes. This publication contributes to the existing body of comprehensive data, with the main difference being that the bulk of this data has exceptionally regular cadence, and has been acquired with a single facility: the Las Cumbres Observatory Global Telescope, a network of robotic telescopes with imaging and spectroscopic capabilities. Only with the community's compilation of time series data is the true nature of SN\,2009ip coming to be understood. High cadence multi-band and spectral coverage has traditionally been rare and challenging to accumulate, but the LCOGT Network can automatically monitor astronomical phenomena on almost any desired timescale; in the future, this can be done without human intervention. In particular, a distributed network like LCOGT mitigates losses due to weather and facilitates the acquisition of daily monitoring. LCOGT will continue to monitor SN\,2009ip as it declines towards its fate.

\acknowledgments
We thank Jon Mauerhan for access to the photometry presented in JM13, and discussion about the late time spectra including the tip about the red-side asymmetry. We thank Andrea Pastorello for access to the spectra presented in AP13 and photometry in Pastorello et al. (2002).

This research has made use of the LCOGT Archive, which is operated by the California Institute of Technology, under contract with the Las Cumbres Observatory.

This research is based on observations obtained at the Southern Astrophysical Research (SOAR) telescope, which is a joint project of the Minist\'{e}rio da Ci\^{e}ncia, Tecnologia, e Inova\c{c}\~{a}o (MCTI) da Rep\'{u}blica Federativa do Brasil, the U.S. National Optical Astronomy Observatory (NOAO), the University of North Carolina at Chapel Hill (UNC), and Michigan State University (MSU). 

This research is based on observations obtained at the Gemini Observatory (PI Howell, program GS-2012A-Q-62), which is operated by the Association of Universities for Research in Astronomy, Inc., under a cooperative agreement with the NSF on behalf of the Gemini partnership: the National Science Foundation (United States), the National Research Council (Canada), CONICYT (Chile), the Australian Research Council (Australia), Minist\'{e}rio da Ci\^{e}ncia, Tecnologia e Inova\c{c}\~{a}o (Brazil) and Ministerio de Ciencia, Tecnolog\'{i}a e Innovaci\'{o}n Productiva (Argentina). 

This research is based on observations obtained at Kitt Peak National Observatory, National Optical Astronomy Observatory, which is operated by the Association of Universities for Research in Astronomy (AURA) under cooperative agreement with the National Science Foundation.

This work has made use of the Weizmann Interactive Supernova Data Repository (WISEREP) at www.weizmann.ac.il/astrophysics/wiserep, and the SUSPECT database at http://nhn.nhn.ou.edu/$\sim$suspect.


{\it Facilities:} \facility{LCOGT, SOAR, KPNO, Gemini}.

\bibliographystyle{apj}
\bibliography{apj-jour,mybib}

\input{tab_allphot.tex}
\input{tab_spec.tex}

\end{document}

%% file: tab_allphot.tex
\begin{center}
\begin{deluxetable*}{ccccccccccccc}
\tablecolumns{13}
\tablecaption{LCOGT photometry for the 2012-B event of SN\,2009ip. \label{tab:allphot}}
\tablehead{\colhead{Date} & \colhead{FTS $g$} & \colhead{FTS $r$} & \colhead{FTS $i$} & \colhead{1m $g$} & \colhead{1m $r$} & \colhead{1m $i$} & \colhead{1m $z$-short} & \colhead{1m $U$} & \colhead{1m $B$} & \colhead{1m $V$} & \colhead{1m $R$} & \colhead{1m $I$}}
\startdata
2012-09-22 & 17.91$\pm$0.10 & 17.56$\pm$0.09 & 17.88$\pm$0.03 & \textellipsis & \textellipsis & \textellipsis & \textellipsis &\textellipsis & \textellipsis & \textellipsis & \textellipsis & \textellipsis \\
2012-09-23 & 17.91$\pm$0.07 & 17.56$\pm$0.06 & 17.89$\pm$0.03 & \textellipsis & \textellipsis & \textellipsis & \textellipsis &\textellipsis & \textellipsis & \textellipsis & \textellipsis & \textellipsis \\
2012-09-24 & 17.10$\pm$0.08 & 16.86$\pm$0.07 & 17.11$\pm$0.02 & \textellipsis & \textellipsis & \textellipsis & \textellipsis &\textellipsis & \textellipsis & \textellipsis & \textellipsis & \textellipsis \\
2012-09-25 &  \textellipsis & 15.22$\pm$0.03 & 15.46$\pm$0.03 & \textellipsis & \textellipsis & \textellipsis & \textellipsis &\textellipsis & \textellipsis & \textellipsis & \textellipsis & \textellipsis \\
2012-09-26 &  \textellipsis & 14.59$\pm$0.10 & 14.83$\pm$0.03 & \textellipsis & \textellipsis & \textellipsis & \textellipsis &\textellipsis & \textellipsis & \textellipsis & \textellipsis & \textellipsis \\
2012-09-27 &  \textellipsis & 14.35$\pm$0.25 & 14.61$\pm$0.04 & \textellipsis & \textellipsis & \textellipsis & \textellipsis &\textellipsis & \textellipsis & \textellipsis & \textellipsis & \textellipsis \\
2012-09-30 & 13.89$\pm$0.16 & 14.01$\pm$0.15 & 14.29$\pm$0.03 & \textellipsis & \textellipsis & \textellipsis & \textellipsis &\textellipsis & \textellipsis & \textellipsis & \textellipsis & \textellipsis \\
2012-10-02 & 13.82$\pm$0.10 & 13.91$\pm$0.09 & 14.19$\pm$0.03 & \textellipsis & \textellipsis & \textellipsis & \textellipsis &\textellipsis & \textellipsis & \textellipsis & \textellipsis & \textellipsis \\
2012-10-03 & 13.73$\pm$0.05 & 13.86$\pm$0.02 & 14.13$\pm$0.03 & \textellipsis & \textellipsis & \textellipsis & \textellipsis &\textellipsis & \textellipsis & \textellipsis & \textellipsis & \textellipsis \\
2012-10-04 & 13.65$\pm$0.06 & 13.79$\pm$0.05 & 14.05$\pm$0.03 & \textellipsis & \textellipsis & \textellipsis & \textellipsis &\textellipsis & \textellipsis & \textellipsis & \textellipsis & \textellipsis \\
2012-10-05 &  \textellipsis &  \textellipsis & 14.04$\pm$0.05 & \textellipsis & \textellipsis & \textellipsis & \textellipsis &\textellipsis & \textellipsis & \textellipsis & \textellipsis & \textellipsis \\
2012-10-06 & 13.60$\pm$0.02 & 13.72$\pm$0.02 & 14.00$\pm$0.01 & \textellipsis & \textellipsis & \textellipsis & \textellipsis &\textellipsis & \textellipsis & \textellipsis & \textellipsis & \textellipsis \\
2012-10-08 & 13.68$\pm$0.08 & 13.76$\pm$0.06 & 14.01$\pm$0.03 & \textellipsis & \textellipsis & \textellipsis & \textellipsis &\textellipsis & \textellipsis & \textellipsis & \textellipsis & \textellipsis \\
2012-10-09 & 13.72$\pm$0.07 & 13.77$\pm$0.06 & 14.01$\pm$0.04 & \textellipsis & \textellipsis & \textellipsis & \textellipsis &\textellipsis & \textellipsis & \textellipsis & \textellipsis & \textellipsis \\
2012-10-10 & 13.74$\pm$0.08 & 13.79$\pm$0.08 & 14.02$\pm$0.03 & \textellipsis & \textellipsis & \textellipsis & \textellipsis &\textellipsis & \textellipsis & \textellipsis & \textellipsis & \textellipsis \\
2012-10-11 & 13.76$\pm$0.03 & 13.80$\pm$0.02 & 14.04$\pm$0.03 & \textellipsis & \textellipsis & \textellipsis & \textellipsis &\textellipsis & \textellipsis & \textellipsis & \textellipsis & \textellipsis \\
2012-10-12 & 13.74$\pm$0.06 & 13.78$\pm$0.06 & 14.01$\pm$0.03 & \textellipsis & \textellipsis & \textellipsis & \textellipsis &\textellipsis & \textellipsis & \textellipsis & \textellipsis & \textellipsis \\
2012-10-13 & 13.73$\pm$0.07 & 13.77$\pm$0.06 & 14.00$\pm$0.03 & \textellipsis & \textellipsis & \textellipsis & \textellipsis &\textellipsis & \textellipsis & \textellipsis & \textellipsis & \textellipsis \\
2012-10-14 & 13.77$\pm$0.07 & 13.79$\pm$0.05 & 14.00$\pm$0.03 & \textellipsis & \textellipsis & \textellipsis & \textellipsis &\textellipsis & \textellipsis & \textellipsis & \textellipsis & \textellipsis \\
2012-10-15 & 13.83$\pm$0.07 & 13.83$\pm$0.06 & 14.05$\pm$0.04 & \textellipsis & \textellipsis & \textellipsis & \textellipsis &\textellipsis & \textellipsis & \textellipsis & \textellipsis & \textellipsis \\
2012-10-17 & 13.99$\pm$0.07 & 13.96$\pm$0.06 & 14.15$\pm$0.03 & \textellipsis & \textellipsis & \textellipsis & \textellipsis &\textellipsis & \textellipsis & \textellipsis & \textellipsis & \textellipsis \\
2012-10-18 & 14.06$\pm$0.21 & 14.01$\pm$0.23 & 14.22$\pm$0.03 & \textellipsis & \textellipsis & \textellipsis & \textellipsis &\textellipsis & \textellipsis & \textellipsis & \textellipsis & \textellipsis \\
2012-10-19 & 14.14$\pm$0.06 & 14.09$\pm$0.06 & 14.28$\pm$0.03 & \textellipsis & \textellipsis & \textellipsis & \textellipsis &\textellipsis & \textellipsis & \textellipsis & \textellipsis & \textellipsis \\
2012-10-20 & 14.18$\pm$0.08 & 14.13$\pm$0.07 & 14.32$\pm$0.03 & \textellipsis & \textellipsis & \textellipsis & \textellipsis &\textellipsis & \textellipsis & \textellipsis & \textellipsis & \textellipsis \\
2012-10-21 &  \textellipsis &  \textellipsis & 14.34$\pm$0.03 & 14.20$\pm$0.07 & 14.16$\pm$0.05 & 14.31$\pm$0.05 & 14.36$\pm$0.05 &  \textellipsis &  \textellipsis &  \textellipsis &  \textellipsis &  \textellipsis \\ 
2012-10-22 & 14.30$\pm$0.03 & 14.20$\pm$0.03 & 14.40$\pm$0.01 & \textellipsis & \textellipsis & \textellipsis & \textellipsis &\textellipsis & \textellipsis & \textellipsis & \textellipsis & \textellipsis \\
2012-10-23 & 14.34$\pm$0.09 & 14.25$\pm$0.06 & 14.43$\pm$0.03 & \textellipsis & \textellipsis & \textellipsis & \textellipsis &\textellipsis & \textellipsis & \textellipsis & \textellipsis & \textellipsis \\
2012-10-24 & 14.41$\pm$0.11 & 14.32$\pm$0.09 & 14.49$\pm$0.03 & \textellipsis & \textellipsis & \textellipsis & \textellipsis &\textellipsis & \textellipsis & \textellipsis & \textellipsis & \textellipsis \\
2012-10-25 & 14.51$\pm$0.12 & 14.39$\pm$0.10 & 14.57$\pm$0.03 & 14.55$\pm$0.15 & 14.42$\pm$0.09 & 14.54$\pm$0.07 & 14.66$\pm$0.11 &  \textellipsis &  \textellipsis &  \textellipsis & 14.23$\pm$0.12 &  \textellipsis \\ 
2012-10-26 &  \textellipsis &  \textellipsis &  \textellipsis &  \textellipsis &  \textellipsis &  \textellipsis & 15.09$\pm$0.86 &  \textellipsis &  \textellipsis &  \textellipsis &  \textellipsis &  \textellipsis \\ 
2012-10-27 &  \textellipsis &  \textellipsis & 14.85$\pm$0.27 & \textellipsis & \textellipsis & \textellipsis & \textellipsis &\textellipsis & \textellipsis & \textellipsis & \textellipsis & \textellipsis \\
2012-10-28 &  \textellipsis &  \textellipsis & 14.95$\pm$0.04 & \textellipsis & \textellipsis & \textellipsis & \textellipsis &\textellipsis & \textellipsis & \textellipsis & \textellipsis & \textellipsis \\
2012-10-29 & \textellipsis & \textellipsis & \textellipsis &14.93$\pm$0.09 & 14.77$\pm$0.07 & 14.89$\pm$0.15 & 14.94$\pm$0.07 &  \textellipsis & 15.32$\pm$0.07 & 14.92$\pm$0.06 & 14.60$\pm$0.10 & 14.48$\pm$0.05 \\ 
2012-10-30 & 15.12$\pm$0.08 & 14.90$\pm$0.07 & 15.07$\pm$0.03 & 15.06$\pm$0.09 & 14.87$\pm$0.05 & 15.03$\pm$0.19 & 15.02$\pm$0.11 &  \textellipsis & 15.42$\pm$0.04 & 15.04$\pm$0.08 & 15.04$\pm$0.73 & 14.58$\pm$0.10 \\ 
2012-10-31 & 15.07$\pm$0.08 & 14.86$\pm$0.08 & 15.06$\pm$0.03 & \textellipsis & \textellipsis & \textellipsis & \textellipsis &\textellipsis & \textellipsis & \textellipsis & \textellipsis & \textellipsis \\
2012-11-02 & 14.88$\pm$0.08 & 14.75$\pm$0.07 & 14.97$\pm$0.01 & 14.97$\pm$0.10 & 14.83$\pm$0.06 & 14.94$\pm$0.11 & 14.93$\pm$0.05 & 15.00$\pm$0.14 & 15.28$\pm$0.03 & 14.90$\pm$0.03 & 14.60$\pm$0.03 & 14.49$\pm$0.05 \\ 
2012-11-03 & 14.82$\pm$0.11 & 14.71$\pm$0.09 & 14.91$\pm$0.01 & 14.89$\pm$0.10 & 14.78$\pm$0.06 & 14.96$\pm$0.22 & 14.88$\pm$0.05 & 14.93$\pm$0.18 & 15.18$\pm$0.05 & 14.84$\pm$0.14 & 14.54$\pm$0.06 & 14.44$\pm$0.09 \\ 
2012-11-04 & 14.79$\pm$0.06 & 14.68$\pm$0.06 & 14.86$\pm$0.03 & 14.86$\pm$0.15 & 14.70$\pm$0.06 & 14.83$\pm$0.14 & 14.88$\pm$0.20 & 14.83$\pm$0.12 & 15.13$\pm$0.05 & 14.83$\pm$0.12 & 14.52$\pm$0.04 & 14.39$\pm$0.09 \\ 
2012-11-05 & 14.87$\pm$0.11 & 14.71$\pm$0.09 & 14.90$\pm$0.04 & 14.89$\pm$0.14 & 14.77$\pm$0.11 & 14.90$\pm$0.13 & 14.88$\pm$0.07 & 15.01$\pm$0.19 & 15.19$\pm$0.09 & 14.86$\pm$0.13 & 14.55$\pm$0.08 & 14.42$\pm$0.12 \\ 
2012-11-06 & 14.94$\pm$0.12 & 14.79$\pm$0.11 & 14.98$\pm$0.09 & 14.95$\pm$0.08 & 14.81$\pm$0.05 & 14.91$\pm$0.13 & 14.88$\pm$0.05 & 15.02$\pm$0.13 & 15.26$\pm$0.06 & 14.90$\pm$0.11 & 14.59$\pm$0.04 & 14.45$\pm$0.08 \\ 
2012-11-07 & \textellipsis & \textellipsis & \textellipsis &15.04$\pm$0.10 & 14.87$\pm$0.04 & 14.98$\pm$0.07 & 15.00$\pm$0.05 &  \textellipsis & 15.38$\pm$0.06 & 14.98$\pm$0.02 & 14.66$\pm$0.04 & 14.52$\pm$0.08 \\ 
2012-11-08 & \textellipsis & \textellipsis & \textellipsis &15.16$\pm$0.12 & 14.94$\pm$0.05 & 15.03$\pm$0.14 & 15.08$\pm$0.05 & 15.33$\pm$0.10 & 15.51$\pm$0.06 & 15.10$\pm$0.09 & 14.74$\pm$0.03 & 14.62$\pm$0.07 \\ 
2012-11-09 & 15.32$\pm$0.45 & 15.06$\pm$0.39 & 15.27$\pm$0.07 & 15.46$\pm$0.25 & 15.18$\pm$0.14 & 15.30$\pm$0.08 & 15.26$\pm$0.11 & 15.47$\pm$0.18 & 15.76$\pm$0.14 & 15.34$\pm$0.16 & 15.01$\pm$0.18 & 14.86$\pm$0.19 \\ 
2012-11-10 & 15.47$\pm$0.08 & 15.14$\pm$0.08 & 15.34$\pm$0.03 & 15.61$\pm$0.17 & 15.34$\pm$0.14 & 15.45$\pm$0.15 & 15.38$\pm$0.15 & 15.88$\pm$0.22 & 15.99$\pm$0.17 & 15.51$\pm$0.15 & 15.14$\pm$0.15 & 15.00$\pm$0.21 \\ 
2012-11-11 & 15.54$\pm$0.06 & 15.23$\pm$0.06 & 15.42$\pm$0.03 & \textellipsis & \textellipsis & \textellipsis & \textellipsis &\textellipsis & \textellipsis & \textellipsis & \textellipsis & \textellipsis \\
2012-11-12 & \textellipsis & \textellipsis & \textellipsis &15.72$\pm$0.08 & 15.38$\pm$0.04 & 15.48$\pm$0.10 & 15.46$\pm$0.07 & 16.09$\pm$0.13 & 16.17$\pm$0.06 & 15.59$\pm$0.02 & 15.19$\pm$0.02 & 15.04$\pm$0.05 \\ 
2012-11-13 & \textellipsis & \textellipsis & \textellipsis &15.78$\pm$0.10 & 15.43$\pm$0.04 & 15.54$\pm$0.11 & 15.48$\pm$0.05 & 16.30$\pm$0.10 & 16.24$\pm$0.06 & 15.67$\pm$0.04 & 15.27$\pm$0.05 & 15.14$\pm$0.12 \\ 
2012-11-14 & \textellipsis & \textellipsis & \textellipsis &15.79$\pm$0.10 & 15.52$\pm$0.15 & 15.61$\pm$0.19 & 15.50$\pm$0.10 & 16.30$\pm$0.08 & 16.27$\pm$0.04 & 15.66$\pm$0.13 & 15.23$\pm$0.12 & 15.14$\pm$0.07 \\ 
2012-11-15 & \textellipsis & \textellipsis & \textellipsis &15.80$\pm$0.10 & 15.44$\pm$0.09 & 15.57$\pm$0.17 & 15.51$\pm$0.07 & 16.37$\pm$0.07 & 16.29$\pm$0.02 & 15.69$\pm$0.03 & 15.28$\pm$0.03 & 15.12$\pm$0.05 \\ 
2012-11-16 & \textellipsis & \textellipsis & \textellipsis &15.81$\pm$0.07 & 15.48$\pm$0.05 & 15.58$\pm$0.10 & 15.57$\pm$0.07 & 16.25$\pm$0.13 & 16.29$\pm$0.04 & 15.69$\pm$0.03 & 15.29$\pm$0.05 & 15.15$\pm$0.07 \\ 
2012-11-17 &  \textellipsis &  \textellipsis & 15.68$\pm$0.03 & 15.86$\pm$0.12 & 15.49$\pm$0.08 & 15.61$\pm$0.13 & 15.54$\pm$0.06 & 16.26$\pm$0.03 & 16.34$\pm$0.05 & 15.74$\pm$0.03 & 15.32$\pm$0.02 & 15.15$\pm$0.05 \\ 
2012-11-18 & 15.90$\pm$0.04 & 15.55$\pm$0.04 & 15.68$\pm$0.02 & 15.97$\pm$0.10 & 15.56$\pm$0.07 & 15.67$\pm$0.15 & 15.62$\pm$0.07 & 16.43$\pm$0.11 & 16.46$\pm$0.02 & 15.79$\pm$0.05 & 15.36$\pm$0.05 & 15.18$\pm$0.01 \\ 
2012-11-19 &  \textellipsis &  \textellipsis & 15.76$\pm$0.02 & 16.03$\pm$0.12 & 15.62$\pm$0.09 & 15.71$\pm$0.17 & 15.65$\pm$0.08 & 16.65$\pm$0.10 & 16.57$\pm$0.07 & 15.91$\pm$0.04 & 15.42$\pm$0.07 & 15.24$\pm$0.04 \\ 
2012-11-20 & 16.10$\pm$0.03 & 15.68$\pm$0.02 & 15.79$\pm$0.02 & 16.21$\pm$0.08 & 15.73$\pm$0.05 & 15.77$\pm$0.10 & 15.74$\pm$0.05 & 16.78$\pm$0.13 & 16.70$\pm$0.09 & 15.98$\pm$0.02 & 15.50$\pm$0.04 & 15.32$\pm$0.09 \\ 
2012-11-21 & 16.26$\pm$0.03 & 15.73$\pm$0.02 & 15.79$\pm$0.02 & 16.26$\pm$0.08 & 15.73$\pm$0.07 & 15.77$\pm$0.15 & 15.76$\pm$0.06 &  \textellipsis & 16.78$\pm$0.05 & 16.01$\pm$0.10 & 15.54$\pm$0.03 & 15.34$\pm$0.07 \\ 
2012-11-22 & 16.75$\pm$0.46 &  \textellipsis & 15.90$\pm$0.19 & 16.30$\pm$0.07 & 15.78$\pm$0.08 & 15.83$\pm$0.11 & 15.75$\pm$0.07 & 16.93$\pm$0.08 & 16.86$\pm$0.05 & 16.09$\pm$0.03 & 15.58$\pm$0.03 & 15.36$\pm$0.08 \\ 
2012-11-23 & \textellipsis & \textellipsis & \textellipsis &16.39$\pm$0.08 & 15.85$\pm$0.05 & 15.87$\pm$0.11 & 15.86$\pm$0.05 & 16.97$\pm$0.14 & 16.90$\pm$0.08 & 16.15$\pm$0.06 & 15.65$\pm$0.07 & 15.40$\pm$0.11 \\ 
2012-11-24 &  \textellipsis &  \textellipsis & 15.92$\pm$0.02 & 16.42$\pm$0.11 & 15.87$\pm$0.05 & 15.87$\pm$0.13 & 15.79$\pm$0.07 & 17.00$\pm$0.15 & 16.92$\pm$0.05 & 16.16$\pm$0.03 & 15.66$\pm$0.04 & 15.41$\pm$0.10 \\ 
2012-11-25 &  \textellipsis &  \textellipsis & 15.92$\pm$0.02 & 16.43$\pm$0.10 & 15.86$\pm$0.06 & 15.89$\pm$0.10 & 15.80$\pm$0.07 & 17.13$\pm$0.12 & 16.97$\pm$0.13 & 16.20$\pm$0.05 & 15.68$\pm$0.05 & 15.42$\pm$0.11 \\ 
2012-11-26 & \textellipsis & \textellipsis & \textellipsis &16.45$\pm$0.06 & 15.89$\pm$0.05 & 15.88$\pm$0.13 & 15.84$\pm$0.06 &  \textellipsis & 17.08$\pm$0.09 & 16.20$\pm$0.11 & 15.66$\pm$0.08 & 15.43$\pm$0.08 \\ 
2012-11-27 & \textellipsis & \textellipsis & \textellipsis &16.50$\pm$0.06 & 15.92$\pm$0.04 & 15.91$\pm$0.13 & 15.84$\pm$0.05 & 17.21$\pm$0.14 & 17.06$\pm$0.07 & 16.25$\pm$0.04 & 15.72$\pm$0.04 & 15.45$\pm$0.08 \\ 
2012-11-29 &  \textellipsis &  \textellipsis &  \textellipsis & 16.57$\pm$0.06 & 15.97$\pm$0.04 & 15.95$\pm$0.06 & 15.88$\pm$0.05 &  \textellipsis & 17.13$\pm$0.08 & 16.32$\pm$0.07 & 15.77$\pm$0.08 & 15.50$\pm$0.14 \\ 
2012-11-30 &  \textellipsis &  \textellipsis &  \textellipsis & 16.64$\pm$0.12 & 16.00$\pm$0.05 & 15.96$\pm$0.07 & 15.95$\pm$0.07 & 17.30$\pm$0.17 & 17.15$\pm$0.08 & 16.33$\pm$0.05 & 15.77$\pm$0.06 & 15.50$\pm$0.11 \\ 
2012-12-01 & 16.58$\pm$0.09 & 15.95$\pm$0.08 & 15.98$\pm$0.03 & \textellipsis & \textellipsis & \textellipsis & \textellipsis &\textellipsis & \textellipsis & \textellipsis & \textellipsis & \textellipsis \\
2012-12-02 & 16.62$\pm$0.12 & 15.94$\pm$0.10 & 15.99$\pm$0.03 & \textellipsis & \textellipsis & \textellipsis & \textellipsis &\textellipsis & \textellipsis & \textellipsis & \textellipsis & \textellipsis \\
2012-12-03 & \textellipsis & \textellipsis & \textellipsis &16.71$\pm$0.07 & 16.04$\pm$0.04 & 16.01$\pm$0.12 & 15.89$\pm$0.06 &  \textellipsis &  \textellipsis &  \textellipsis &  \textellipsis &  \textellipsis \\ 
2012-12-04 & 16.74$\pm$0.08 & 16.04$\pm$0.07 & 16.07$\pm$0.03 & 16.86$\pm$0.07 & 16.14$\pm$0.05 & 16.10$\pm$0.08 & 16.01$\pm$0.10 &  \textellipsis &  \textellipsis &  \textellipsis &  \textellipsis &  \textellipsis \\ 
2012-12-05 & 16.93$\pm$0.11 & 16.12$\pm$0.10 & 16.12$\pm$0.03 & \textellipsis & \textellipsis & \textellipsis & \textellipsis &\textellipsis & \textellipsis & \textellipsis & \textellipsis & \textellipsis \\
2012-12-06 & 17.07$\pm$0.10 & 16.20$\pm$0.08 & 16.21$\pm$0.03 & 17.19$\pm$0.06 & 16.28$\pm$0.05 & 16.20$\pm$0.11 & 16.13$\pm$0.06 &  \textellipsis &  \textellipsis &  \textellipsis &  \textellipsis &  \textellipsis \\ 
2012-12-07 & 17.27$\pm$0.14 & 16.33$\pm$0.13 & 16.28$\pm$0.03 & 17.36$\pm$0.09 & 16.41$\pm$0.06 & 16.31$\pm$0.08 & 16.20$\pm$0.08 &  \textellipsis &  \textellipsis &  \textellipsis &  \textellipsis &  \textellipsis \\ 
2012-12-08 & 17.37$\pm$0.08 & 16.35$\pm$0.07 & 16.37$\pm$0.03 &  \textellipsis &  \textellipsis &  \textellipsis &  \textellipsis &  \textellipsis &  \textellipsis &  \textellipsis &  \textellipsis &  \textellipsis \\ 
2012-12-10 & \textellipsis & \textellipsis & \textellipsis &17.71$\pm$0.09 & 16.68$\pm$0.05 & 16.62$\pm$0.08 & 16.46$\pm$0.07 &  \textellipsis &  \textellipsis &  \textellipsis &  \textellipsis &  \textellipsis \\ 
2012-12-11 & 17.70$\pm$0.09 &  \textellipsis &  \textellipsis & \textellipsis & \textellipsis & \textellipsis & \textellipsis &\textellipsis & \textellipsis & \textellipsis & \textellipsis & \textellipsis \\
2012-12-12 & 17.97$\pm$0.10 & 16.86$\pm$0.10 & 17.02$\pm$0.03 & 18.15$\pm$0.07 & 17.04$\pm$0.05 & 17.13$\pm$0.08 & 16.84$\pm$0.14 &  \textellipsis &  \textellipsis &  \textellipsis &  \textellipsis &  \textellipsis \\ 
2012-12-13 & 18.20$\pm$0.07 & 17.04$\pm$0.05 & 17.26$\pm$0.03 & 18.30$\pm$0.09 & 17.21$\pm$0.06 & 17.31$\pm$0.09 & 17.05$\pm$0.15 &  \textellipsis &  \textellipsis &  \textellipsis &  \textellipsis &  \textellipsis \\ 
2012-12-14 & 18.31$\pm$0.13 & 17.24$\pm$0.11 &  \textellipsis & 18.45$\pm$0.11 & 17.35$\pm$0.08 & 17.54$\pm$0.07 & 17.30$\pm$0.10 &  \textellipsis &  \textellipsis &  \textellipsis &  \textellipsis &  \textellipsis \\ 
2012-12-15 & \textellipsis & \textellipsis & \textellipsis &18.52$\pm$0.13 & 17.41$\pm$0.08 & 17.59$\pm$0.08 & 17.19$\pm$0.10 &  \textellipsis &  \textellipsis &  \textellipsis &  \textellipsis &  \textellipsis \\ 
2012-12-16 & \textellipsis & \textellipsis & \textellipsis &18.60$\pm$0.08 & 17.45$\pm$0.06 & 17.67$\pm$0.09 & 17.62$\pm$0.13 &  \textellipsis &  \textellipsis &  \textellipsis &  \textellipsis &  \textellipsis \\ 
2012-12-17 & 18.59$\pm$0.12 & 17.32$\pm$0.10 & 17.66$\pm$0.03 & 18.52$\pm$0.12 & 17.41$\pm$0.06 & 17.61$\pm$0.12 &  \textellipsis &  \textellipsis &  \textellipsis &  \textellipsis &  \textellipsis &  \textellipsis \\ 
2012-12-23 & 18.60$\pm$0.11 & 17.41$\pm$0.10 & 17.75$\pm$0.03 & \textellipsis & \textellipsis & \textellipsis & \textellipsis &\textellipsis & \textellipsis & \textellipsis & \textellipsis & \textellipsis \\
2012-12-24 &  \textellipsis & 17.41$\pm$0.10 & 17.76$\pm$0.05 & \textellipsis & \textellipsis & \textellipsis & \textellipsis &\textellipsis & \textellipsis & \textellipsis & \textellipsis & \textellipsis \\
2012-12-29 &  \textellipsis & 17.42$\pm$0.04 & 17.82$\pm$0.04 & \textellipsis & \textellipsis & \textellipsis & \textellipsis &\textellipsis & \textellipsis & \textellipsis & \textellipsis & \textellipsis \\
2012-12-31 &  \textellipsis & 17.50$\pm$0.05 & 17.86$\pm$0.05 & \textellipsis & \textellipsis & \textellipsis & \textellipsis &\textellipsis & \textellipsis & \textellipsis & \textellipsis & \textellipsis \\
2013-04-02 & \textellipsis & \textellipsis & \textellipsis &19.62$\pm$0.14 & 18.16$\pm$0.06 &  \textellipsis &  \textellipsis &  \textellipsis &  \textellipsis &  \textellipsis &  \textellipsis &  \textellipsis \\ 
2013-04-04 & \textellipsis & \textellipsis & \textellipsis &19.59$\pm$0.08 & 18.18$\pm$0.08 &  \textellipsis &  \textellipsis &  \textellipsis &  \textellipsis &  \textellipsis &  \textellipsis &  \textellipsis \\ 
2013-04-05 & \textellipsis & \textellipsis & \textellipsis & \textellipsis & 18.19$\pm$0.08 & 18.89$\pm$0.14 &  \textellipsis &  \textellipsis &  \textellipsis &  \textellipsis &  \textellipsis &  \textellipsis \\ 
2013-04-06 & \textellipsis & \textellipsis & \textellipsis &19.53$\pm$0.13 & 18.16$\pm$0.07 &  \textellipsis &  \textellipsis &  \textellipsis &  \textellipsis &  \textellipsis &  \textellipsis &  \textellipsis \\ 
2013-04-10 & \textellipsis & \textellipsis & \textellipsis & \textellipsis & 18.19$\pm$0.08 & 18.85$\pm$0.17 &  \textellipsis &  \textellipsis &  \textellipsis &  \textellipsis &  \textellipsis &  \textellipsis \\ 
2013-04-11 & \textellipsis & \textellipsis & \textellipsis &19.64$\pm$0.14 & 18.19$\pm$0.07 & 18.86$\pm$0.09 &  \textellipsis &  \textellipsis &  \textellipsis &  \textellipsis &  \textellipsis &  \textellipsis \\ 
2013-04-12 & \textellipsis & \textellipsis & \textellipsis &19.59$\pm$0.09 & 18.18$\pm$0.07 & 18.71$\pm$0.13 &  \textellipsis &  \textellipsis &  \textellipsis &  \textellipsis &  \textellipsis &  \textellipsis \\ 
2013-04-16 & \textellipsis & \textellipsis & \textellipsis &19.57$\pm$0.11 & 18.18$\pm$0.09 & 18.91$\pm$0.08 &  \textellipsis &  \textellipsis &  \textellipsis &  \textellipsis &  \textellipsis &  \textellipsis \\ 
2013-04-17 & \textellipsis & \textellipsis & \textellipsis &19.56$\pm$0.12 & 18.22$\pm$0.06 & 18.96$\pm$0.08 &  \textellipsis &  \textellipsis &  \textellipsis &  \textellipsis &  \textellipsis &  \textellipsis \\ 
2013-04-18 & \textellipsis & \textellipsis & \textellipsis &19.56$\pm$0.10 & 18.19$\pm$0.08 &  \textellipsis &  \textellipsis &  \textellipsis &  \textellipsis &  \textellipsis &  \textellipsis &  \textellipsis \\ 
2013-04-19 & \textellipsis & \textellipsis & \textellipsis & \textellipsis & 18.23$\pm$0.06 &  \textellipsis &  \textellipsis &  \textellipsis &  \textellipsis &  \textellipsis &  \textellipsis &  \textellipsis \\ 
2013-04-20 & \textellipsis & \textellipsis & \textellipsis &19.61$\pm$0.11 & 18.21$\pm$0.05 & 18.86$\pm$0.09 &  \textellipsis &  \textellipsis &  \textellipsis &  \textellipsis &  \textellipsis &  \textellipsis \\ 
2013-04-21 & \textellipsis & \textellipsis & \textellipsis & \textellipsis & 18.25$\pm$0.08 &  \textellipsis &  \textellipsis &  \textellipsis &  \textellipsis &  \textellipsis &  \textellipsis &  \textellipsis \\ 
2013-04-22 & \textellipsis & \textellipsis & \textellipsis &19.69$\pm$0.08 & 18.13$\pm$0.13 &  \textellipsis &  \textellipsis &  \textellipsis &  \textellipsis &  \textellipsis &  \textellipsis &  \textellipsis \\ 
2013-04-24 & \textellipsis & \textellipsis & \textellipsis &19.57$\pm$0.15 & 18.23$\pm$0.11 & 19.04$\pm$0.10 &  \textellipsis &  \textellipsis &  \textellipsis &  \textellipsis &  \textellipsis &  \textellipsis \\ 
2013-04-25 & \textellipsis & \textellipsis & \textellipsis & \textellipsis & 18.29$\pm$0.10 & 18.90$\pm$0.17 &  \textellipsis &  \textellipsis &  \textellipsis &  \textellipsis &  \textellipsis &  \textellipsis \\ 
2013-04-26 & \textellipsis & \textellipsis & \textellipsis & \textellipsis & 18.26$\pm$0.13 &  \textellipsis &  \textellipsis &  \textellipsis &  \textellipsis &  \textellipsis &  \textellipsis &  \textellipsis \\ 
2013-04-28 & \textellipsis & \textellipsis & \textellipsis &19.78$\pm$0.17 & 18.32$\pm$0.07 & 19.03$\pm$0.11 &  \textellipsis &  \textellipsis &  \textellipsis &  \textellipsis &  \textellipsis &  \textellipsis \\ 
2013-04-29 & \textellipsis & \textellipsis & \textellipsis &19.57$\pm$0.14 & 18.31$\pm$0.13 & 18.92$\pm$0.14 &  \textellipsis &  \textellipsis &  \textellipsis &  \textellipsis &  \textellipsis &  \textellipsis \\ 
2013-05-02 & \textellipsis & \textellipsis & \textellipsis &19.81$\pm$0.22 & 18.37$\pm$0.16 & 19.07$\pm$0.26 &  \textellipsis &  \textellipsis & 19.90$\pm$0.13 & 19.48$\pm$0.27 &  \textellipsis &  \textellipsis \\ 
2013-05-03 & \textellipsis & \textellipsis & \textellipsis &19.70$\pm$0.12 & 18.41$\pm$0.07 & 19.10$\pm$0.09 &  \textellipsis &  \textellipsis & 20.24$\pm$0.13 & 19.67$\pm$0.28 &  \textellipsis &  \textellipsis \\ 
2013-05-05 & \textellipsis & \textellipsis & \textellipsis &19.84$\pm$0.11 & 18.35$\pm$0.08 & 19.17$\pm$0.10 &  \textellipsis &  \textellipsis & 20.34$\pm$0.21 & 19.71$\pm$0.11 &  \textellipsis &  \textellipsis \\ 
2013-05-06 & \textellipsis & \textellipsis & \textellipsis &19.70$\pm$0.14 &  \textellipsis & 19.23$\pm$0.06 &  \textellipsis &  \textellipsis &  \textellipsis &  \textellipsis &  \textellipsis &  \textellipsis \\ 
2013-05-08 & \textellipsis & \textellipsis & \textellipsis &19.87$\pm$0.15 & 18.33$\pm$0.06 & 19.25$\pm$0.13 &  \textellipsis &  \textellipsis & 20.23$\pm$0.26 & 19.62$\pm$0.17 &  \textellipsis &  \textellipsis \\ 
2013-05-10 & \textellipsis & \textellipsis & \textellipsis &19.86$\pm$0.15 & 18.42$\pm$0.07 & 19.26$\pm$0.07 &  \textellipsis &  \textellipsis & 20.23$\pm$0.13 & 19.69$\pm$0.17 &  \textellipsis &  \textellipsis \\ 
2013-05-15 & \textellipsis & \textellipsis & \textellipsis &19.99$\pm$0.12 & 18.45$\pm$0.09 & 19.35$\pm$0.09 &  \textellipsis &  \textellipsis & 20.27$\pm$0.12 & 19.74$\pm$0.08 &  \textellipsis &  \textellipsis \\ 
2013-06-02 & \textellipsis & \textellipsis & \textellipsis &20.18$\pm$0.11 & 18.63$\pm$0.07 & 19.44$\pm$0.08 &  \textellipsis &  \textellipsis & 20.80$\pm$0.13 & 19.87$\pm$0.11 &  \textellipsis &  \textellipsis \\ 
2013-06-03 & \textellipsis & \textellipsis & \textellipsis &20.17$\pm$0.12 & 18.61$\pm$0.16 & 19.45$\pm$0.08 &  \textellipsis &  \textellipsis & 20.47$\pm$0.12 & 19.92$\pm$0.17 &  \textellipsis &  \textellipsis \\ 
2013-06-04 & \textellipsis & \textellipsis & \textellipsis &20.22$\pm$0.13 & 18.64$\pm$0.07 & 19.55$\pm$0.12 &  \textellipsis &  \textellipsis & 20.46$\pm$0.11 & 19.88$\pm$0.13 &  \textellipsis &  \textellipsis \\ 
2013-06-05 & \textellipsis & \textellipsis & \textellipsis & \textellipsis &  \textellipsis &  \textellipsis &  \textellipsis &  \textellipsis & 20.61$\pm$0.19 & 20.01$\pm$0.14 &  \textellipsis &  \textellipsis \\ 
2013-06-06 & \textellipsis & \textellipsis & \textellipsis &20.16$\pm$0.13 & 18.66$\pm$0.07 & 19.52$\pm$0.09 &  \textellipsis &  \textellipsis & 20.40$\pm$0.12 &  \textellipsis &  \textellipsis &  \textellipsis \\ 
2013-06-08 & \textellipsis & \textellipsis & \textellipsis &20.21$\pm$0.14 & 18.66$\pm$0.06 & 19.67$\pm$0.08 &  \textellipsis &  \textellipsis &  \textellipsis &  \textellipsis &  \textellipsis &  \textellipsis \\ 
2013-06-09 & \textellipsis & \textellipsis & \textellipsis & \textellipsis & 18.42$\pm$0.55 &  \textellipsis &  \textellipsis &  \textellipsis &  \textellipsis &  \textellipsis &  \textellipsis &  \textellipsis \\ 
2013-06-11 & \textellipsis & \textellipsis & \textellipsis &20.19$\pm$0.13 & 18.65$\pm$0.07 & 19.53$\pm$0.09 &  \textellipsis &  \textellipsis & 20.41$\pm$0.11 & 20.05$\pm$0.19 &  \textellipsis &  \textellipsis \\ 
2013-06-13 & \textellipsis & \textellipsis & \textellipsis & \textellipsis &  \textellipsis &  \textellipsis &  \textellipsis &  \textellipsis & 20.35$\pm$0.12 & 20.00$\pm$0.34 &  \textellipsis &  \textellipsis \\ 
2013-06-15 & \textellipsis & \textellipsis & \textellipsis &20.21$\pm$0.11 & 18.64$\pm$0.10 & 19.56$\pm$0.09 &  \textellipsis &  \textellipsis & 20.60$\pm$0.11 & 20.01$\pm$0.17 &  \textellipsis &  \textellipsis \\ 
2013-06-19 & \textellipsis & \textellipsis & \textellipsis &20.16$\pm$0.14 & 18.67$\pm$0.06 &  \textellipsis &  \textellipsis &  \textellipsis &  \textellipsis & 20.11$\pm$0.13 &  \textellipsis &  \textellipsis \\ 
2013-06-23 & \textellipsis & \textellipsis & \textellipsis &20.16$\pm$0.16 & 18.72$\pm$0.08 & 19.72$\pm$0.12 &  \textellipsis &  \textellipsis &  \textellipsis &  \textellipsis &  \textellipsis &  \textellipsis \\ 
2013-06-25 & \textellipsis & \textellipsis & \textellipsis & \textellipsis & 18.65$\pm$0.09 &  \textellipsis &  \textellipsis &  \textellipsis &  \textellipsis &  \textellipsis &  \textellipsis &  \textellipsis \\ 
2013-07-01 & \textellipsis & \textellipsis & \textellipsis &20.32$\pm$0.14 & 18.72$\pm$0.06 & 19.52$\pm$0.13 &  \textellipsis &  \textellipsis & 20.56$\pm$0.26 & 20.14$\pm$0.18 &  \textellipsis &  \textellipsis \\ 
2013-07-02 & \textellipsis & \textellipsis & \textellipsis &20.30$\pm$0.11 & 18.76$\pm$0.07 & 19.63$\pm$0.11 &  \textellipsis &  \textellipsis &  \textellipsis & 20.20$\pm$0.11 &  \textellipsis &  \textellipsis \\ 
2013-07-06 & \textellipsis & \textellipsis & \textellipsis &20.22$\pm$0.25 & 18.71$\pm$0.21 & 19.60$\pm$0.15 &  \textellipsis &  \textellipsis & 20.73$\pm$0.29 & 20.06$\pm$0.33 &  \textellipsis &  \textellipsis \\ 
2013-07-08 & \textellipsis & \textellipsis & \textellipsis &20.24$\pm$0.17 & 18.75$\pm$0.07 & 19.60$\pm$0.13 &  \textellipsis &  \textellipsis & 20.68$\pm$0.21 & 20.12$\pm$0.20 &  \textellipsis &  \textellipsis \\ 
2013-07-10 & \textellipsis & \textellipsis & \textellipsis &20.18$\pm$0.13 & 18.69$\pm$0.09 & 19.51$\pm$0.10 &  \textellipsis &  \textellipsis & 20.71$\pm$0.17 & 20.16$\pm$0.24 &  \textellipsis &  \textellipsis \\ 
2013-07-12 & \textellipsis & \textellipsis & \textellipsis &20.28$\pm$0.11 & 18.74$\pm$0.06 & 19.64$\pm$0.10 &  \textellipsis &  \textellipsis & 20.39$\pm$0.11 & 20.13$\pm$0.18 &  \textellipsis &  \textellipsis \\ 
2013-07-13 & \textellipsis & \textellipsis & \textellipsis &20.26$\pm$0.15 & 18.74$\pm$0.27 & 19.46$\pm$0.13 &  \textellipsis &  \textellipsis & 20.63$\pm$0.14 & 20.05$\pm$0.13 &  \textellipsis &  \textellipsis \\ 
2013-07-14 & \textellipsis & \textellipsis & \textellipsis &20.26$\pm$0.11 & 18.82$\pm$0.07 & 19.66$\pm$0.07 &  \textellipsis &  \textellipsis & 20.64$\pm$0.13 & 20.08$\pm$0.17 &  \textellipsis &  \textellipsis \\ 
2013-07-16 & \textellipsis & \textellipsis & \textellipsis &20.29$\pm$0.16 & 18.78$\pm$0.09 & 19.67$\pm$0.12 &  \textellipsis &  \textellipsis & 20.82$\pm$0.14 & 20.16$\pm$0.18 &  \textellipsis &  \textellipsis \\ 
2013-07-17 & \textellipsis & \textellipsis & \textellipsis &20.17$\pm$0.15 & 18.74$\pm$0.07 & 19.68$\pm$0.12 &  \textellipsis &  \textellipsis & 20.46$\pm$0.12 & 20.10$\pm$0.18 &  \textellipsis &  \textellipsis \\ 
2013-07-18 & \textellipsis & \textellipsis & \textellipsis &20.21$\pm$0.11 & 18.78$\pm$0.09 & 19.73$\pm$0.12 &  \textellipsis &  \textellipsis &  \textellipsis & 20.08$\pm$0.11 &  \textellipsis &  \textellipsis \\ 
2013-07-24 & \textellipsis & \textellipsis & \textellipsis & \textellipsis & 18.81$\pm$0.11 &  \textellipsis &  \textellipsis &  \textellipsis &  \textellipsis &  \textellipsis &  \textellipsis &  \textellipsis \\ 
2013-07-25 & \textellipsis & \textellipsis & \textellipsis & \textellipsis & 18.74$\pm$0.09 &  \textellipsis &  \textellipsis &  \textellipsis &  \textellipsis &  \textellipsis &  \textellipsis &  \textellipsis \\ 
2013-07-29 & \textellipsis & \textellipsis & \textellipsis &20.43$\pm$0.15 & 18.79$\pm$0.07 &  \textellipsis &  \textellipsis &  \textellipsis &  \textellipsis &  \textellipsis &  \textellipsis &  \textellipsis \\ 
2013-07-30 & \textellipsis & \textellipsis & \textellipsis &20.32$\pm$0.12 & 18.79$\pm$0.06 & 19.70$\pm$0.11 &  \textellipsis &  \textellipsis & 20.59$\pm$0.15 & 20.05$\pm$0.18 &  \textellipsis &  \textellipsis \\ 
2013-07-31 & \textellipsis & \textellipsis & \textellipsis &20.32$\pm$0.17 & 18.77$\pm$0.08 &  \textellipsis &  \textellipsis &  \textellipsis & 20.68$\pm$0.13 & 20.04$\pm$0.31 &  \textellipsis &  \textellipsis \\ 
2013-08-03 & \textellipsis & \textellipsis & \textellipsis & \textellipsis &  \textellipsis &  \textellipsis &  \textellipsis &  \textellipsis & 20.81$\pm$0.12 & 20.12$\pm$0.19 &  \textellipsis &  \textellipsis \\ 
2013-08-04 & \textellipsis & \textellipsis & \textellipsis &20.14$\pm$0.13 & 18.75$\pm$0.07 & 19.76$\pm$0.10 &  \textellipsis &  \textellipsis &  \textellipsis &  \textellipsis &  \textellipsis &  \textellipsis \\ 
2013-08-08 & \textellipsis & \textellipsis & \textellipsis &20.34$\pm$0.11 & 18.87$\pm$0.09 & 19.79$\pm$0.09 &  \textellipsis &  \textellipsis & 20.48$\pm$0.16 & 20.05$\pm$0.25 &  \textellipsis &  \textellipsis \\ 
2013-08-09 & \textellipsis & \textellipsis & \textellipsis &20.60$\pm$0.15 & 18.94$\pm$0.11 &  \textellipsis &  \textellipsis &  \textellipsis &  \textellipsis &  \textellipsis &  \textellipsis &  \textellipsis \\ 
2013-08-10 & \textellipsis & \textellipsis & \textellipsis & \textellipsis &  \textellipsis &  \textellipsis &  \textellipsis &  \textellipsis & 20.44$\pm$0.12 & 20.13$\pm$0.14 &  \textellipsis &  \textellipsis \\ 
2013-08-12 & \textellipsis & \textellipsis & \textellipsis & \textellipsis & 18.74$\pm$0.05 & 19.77$\pm$0.12 &  \textellipsis &  \textellipsis &  \textellipsis &  \textellipsis &  \textellipsis &  \textellipsis \\ 
2013-08-14 & \textellipsis & \textellipsis & \textellipsis &20.37$\pm$0.14 & 18.89$\pm$0.08 &  \textellipsis &  \textellipsis &  \textellipsis &  \textellipsis & 20.15$\pm$0.12 &  \textellipsis &  \textellipsis \\ 
2013-08-16 & \textellipsis & \textellipsis & \textellipsis &20.31$\pm$0.16 & 18.91$\pm$0.07 &  \textellipsis &  \textellipsis &  \textellipsis &  \textellipsis & 20.01$\pm$0.14 &  \textellipsis &  \textellipsis \\ 
2013-08-20 & \textellipsis & \textellipsis & \textellipsis & \textellipsis & 18.89$\pm$0.13 &  \textellipsis &  \textellipsis &  \textellipsis &  \textellipsis &  \textellipsis &  \textellipsis &  \textellipsis \\ 
2013-08-21 & \textellipsis & \textellipsis & \textellipsis & \textellipsis & 18.87$\pm$0.10 &  \textellipsis &  \textellipsis &  \textellipsis &  \textellipsis &  \textellipsis &  \textellipsis &  \textellipsis \\ 
2013-08-24 & \textellipsis & \textellipsis & \textellipsis &20.14$\pm$0.15 & 18.87$\pm$0.08 &  \textellipsis &  \textellipsis &  \textellipsis &  \textellipsis &  \textellipsis &  \textellipsis &  \textellipsis \\ 
2013-08-26 & \textellipsis & \textellipsis & \textellipsis &20.41$\pm$0.16 & 18.91$\pm$0.05 & 19.84$\pm$0.12 &  \textellipsis &  \textellipsis &  \textellipsis & 20.32$\pm$0.10 &  \textellipsis &  \textellipsis \\ 
2013-08-28 & \textellipsis & \textellipsis & \textellipsis &20.39$\pm$0.13 & 18.92$\pm$0.06 & 19.88$\pm$0.10 &  \textellipsis &  \textellipsis & 20.73$\pm$0.19 &  \textellipsis &  \textellipsis &  \textellipsis \\ 
2013-08-30 & \textellipsis & \textellipsis & \textellipsis &20.35$\pm$0.12 & 18.86$\pm$0.09 &  \textellipsis &  \textellipsis &  \textellipsis &  \textellipsis &  \textellipsis &  \textellipsis &  \textellipsis \\ 
2013-08-31 & \textellipsis & \textellipsis & \textellipsis &20.39$\pm$0.71 & 18.88$\pm$0.54 &  \textellipsis &  \textellipsis &  \textellipsis &  \textellipsis &  \textellipsis &  \textellipsis &  \textellipsis \\ 
2013-09-02 & \textellipsis & \textellipsis & \textellipsis & \textellipsis &  \textellipsis &  \textellipsis &  \textellipsis &  \textellipsis &  \textellipsis & 20.29$\pm$0.19 &  \textellipsis &  \textellipsis \\ 
2013-09-04 & \textellipsis & \textellipsis & \textellipsis & \textellipsis &  \textellipsis &  \textellipsis &  \textellipsis &  \textellipsis &  \textellipsis & 20.37$\pm$0.13 &  \textellipsis &  \textellipsis \\ 
2013-09-05 & \textellipsis & \textellipsis & \textellipsis &20.35$\pm$0.14 & 18.89$\pm$0.10 &  \textellipsis &  \textellipsis &  \textellipsis &  \textellipsis &  \textellipsis &  \textellipsis &  \textellipsis \\ 
2013-09-06 & \textellipsis & \textellipsis & \textellipsis &20.41$\pm$0.13 & 18.95$\pm$0.14 & 20.05$\pm$0.11 &  \textellipsis &  \textellipsis &  \textellipsis &  \textellipsis &  \textellipsis &  \textellipsis \\ 
2013-09-09 & \textellipsis & \textellipsis & \textellipsis &20.45$\pm$0.12 & 18.93$\pm$0.05 & 19.82$\pm$0.12 &  \textellipsis &  \textellipsis & 20.35$\pm$0.13 &  \textellipsis &  \textellipsis &  \textellipsis \\ 
2013-09-11 & \textellipsis & \textellipsis & \textellipsis &20.22$\pm$0.13 & 18.93$\pm$0.11 &  \textellipsis &  \textellipsis &  \textellipsis & 20.52$\pm$0.59 &  \textellipsis &  \textellipsis &  \textellipsis \\ 
2013-09-13 & \textellipsis & \textellipsis & \textellipsis & \textellipsis & 18.93$\pm$0.12 &  \textellipsis &  \textellipsis &  \textellipsis &  \textellipsis &  \textellipsis &  \textellipsis &  \textellipsis \\ 
2013-09-23 & \textellipsis & \textellipsis & \textellipsis &20.44$\pm$0.13 & 19.05$\pm$0.07 & 20.02$\pm$0.19 &  \textellipsis &  \textellipsis &  \textellipsis & 20.54$\pm$0.13 &  \textellipsis &  \textellipsis \\ 
2013-09-27 & \textellipsis & \textellipsis & \textellipsis & \textellipsis &  \textellipsis &  \textellipsis &  \textellipsis &  \textellipsis & 20.84$\pm$0.18 & 20.59$\pm$0.15 &  \textellipsis &  \textellipsis \\ 
2013-09-28 & \textellipsis & \textellipsis & \textellipsis &20.50$\pm$0.11 & 19.08$\pm$0.05 & 20.06$\pm$0.10 &  \textellipsis &  \textellipsis &  \textellipsis &  \textellipsis &  \textellipsis &  \textellipsis \\ 
2013-09-30 & \textellipsis & \textellipsis & \textellipsis &20.37$\pm$0.13 & 18.99$\pm$0.10 & 19.97$\pm$0.12 &  \textellipsis &  \textellipsis &  \textellipsis &  \textellipsis &  \textellipsis &  \textellipsis \\ 
2013-10-07 & \textellipsis & \textellipsis & \textellipsis & \textellipsis & 19.00$\pm$0.12 &  \textellipsis &  \textellipsis &  \textellipsis &  \textellipsis &  \textellipsis &  \textellipsis &  \textellipsis \\ 
2013-10-09 & \textellipsis & \textellipsis & \textellipsis & \textellipsis &  \textellipsis &  \textellipsis &  \textellipsis &  \textellipsis & 21.01$\pm$0.14 &  \textellipsis &  \textellipsis &  \textellipsis \\ 
2013-10-14 & \textellipsis & \textellipsis & \textellipsis & \textellipsis & 19.06$\pm$0.09 &  \textellipsis &  \textellipsis &  \textellipsis &  \textellipsis &  \textellipsis &  \textellipsis &  \textellipsis \\ 
2013-10-19 & \textellipsis & \textellipsis & \textellipsis & \textellipsis & 19.04$\pm$0.07 &  \textellipsis &  \textellipsis &  \textellipsis &  \textellipsis &  \textellipsis &  \textellipsis &  \textellipsis \\ 
2013-10-25 & \textellipsis & \textellipsis & \textellipsis & \textellipsis & 19.04$\pm$0.06 &  \textellipsis &  \textellipsis &  \textellipsis & 20.93$\pm$0.13 &  \textellipsis &  \textellipsis &  \textellipsis \\ 
\enddata
\end{deluxetable*}
\end{center}

%% file: tab_spec.tex
\begin{deluxetable*}{llllll}[h]
\tablecolumns{5}
\tablecaption{List of spectroscopy for SN\,2009ip. \label{tab:spec}}
\tablehead{
\colhead{UT Date} & \colhead{MJ Date} & \colhead{Telescope} & \colhead{Instrument} & \colhead{Phase} & \colhead{Era} \\
}
\startdata
2012-09-22 & 56192 & SOAR          & Goodman & -13 & rise \\
2012-09-25 & 56195 & Faulkes South & Floyds  & -10 & rise \\
2012-09-26 & 56196 & Faulkes South & Floyds  & -9  & rise \\
2012-09-27 & 56197 & Faulkes South & Floyds  & -8  & rise \\
2012-10-01 & 56201 & Faulkes South & Floyds  & -4  & rise \\
2012-10-02 & 56202 & Faulkes South & Floyds  & -3  & rise \\
2012-10-03 & 56203 & Faulkes South & Floyds  & -2  & rise \\
2012-10-04 & 56204 & Faulkes South & Floyds  & -1  & rise \\
2012-10-05 & 56205 & Faulkes South & Floyds  & 0   & rise \\
2012-10-09 & 56209 & Faulkes South & Floyds  & 4   & decline \\
2012-10-11 & 56211 & Mayall        & RC      & 6   & decline \\
2012-10-12 & 56212 & Mayall        & RC      & 7   & decline \\
2012-10-13 & 56213 & Mayall        & RC      & 8   & decline \\
2012-10-14 & 56214 & Mayall        & RC      & 9   & decline \\
2012-10-15 & 56215 & Mayall        & RC      & 10  & decline \\
2012-10-20 & 56220 & Faulkes South & Floyds  & 15  & decline \\
2012-10-30 & 56230 & Faulkes South & Floyds  & 25  & bump \\
2012-11-02 & 56233 & Faulkes South & Floyds  & 27  & bump \\
2012-11-04 & 56235 & Faulkes South & Floyds  & 29  & bump \\
2012-11-10 & 56241 & Faulkes South & Floyds  & 35  & bump \\
2012-11-13 & 56244 & Faulkes South & Floyds  & 38  & bump \\
2012-11-29 & 56260 & Faulkes South & Floyds  & 54  & knee \\
2012-12-03 & 56264 & Faulkes South & Floyds  & 58  & knee \\
2012-12-16 & 56277 & Faulkes South & Floyds  & 71  & ankle \\
2013-06-12 & 56455 & Gemini South  & GMOS-S  & 249 & late
\enddata
\end{deluxetable*}